\documentclass[prb,aps,10pt,superscriptaddress,twocolumn,floatfix]{revtex4-2}
\usepackage[bottom=2cm,right=2cm,left=2cm,top=2.5cm]{geometry}
\usepackage{graphicx}
\usepackage[nice]{nicefrac}
\usepackage{color,xcolor}
\usepackage{amsmath,amssymb,bm,bbm}
\usepackage{amsmath}
\usepackage{enumitem} % allow to adjust line spacing of enumerate
\usepackage{braket}
\usepackage{bbold}
\usepackage{subcaption}
\usepackage[percent]{overpic} % overlay subcaption labels manually
\usepackage{comment}

\usepackage[plainpages=false,pdfpagelabels,colorlinks=true,linkcolor=red,urlcolor=blue,citecolor=blue,pdftitle={},pdfauthor={},pdfdisplaydoctitle=true,pdfduplex=DuplexFlipLongEdge]{hyperref}

\usepackage{xparse} % needed for definition of tensor product sign with sub- & superscripts
\NewDocumentCommand{\tens}{e{_^}}{%
  \mathbin{\mathop{\otimes}\displaylimits
    \IfValueT{#1}{_{#1}}
    \IfValueT{#2}{^{#2}}
  }%
}

\begin{document}

\title{Extension of the adiabatic theorem}

\author{S. Damerow}
\affiliation{Institute for Theoretical Physics, Georg-August-Universität Göttingen, Friedrich-Hund-Platz 1, 37077 Göttingen, Germany}
\author{S. Kehrein}
\affiliation{Institute for Theoretical Physics, Georg-August-Universität Göttingen, Friedrich-Hund-Platz 1, 37077 Göttingen, Germany}

\begin{abstract}
We examine the validity of a potential extension of the adiabatic theorem to quantum quenches, i.e., nonadiabatic changes. In particular, the transverse field Ising model (TFIM) and the axial next nearest neighbor Ising (ANNNI) model are studied. The proposed extension of the adiabatic theorem is stated as follows: Consider the overlap between the initial ground state and the postquench Hamiltonian eigenstates for quenches within the same phase. This overlap is largest for the postquench ground state. In the case of the TFIM, this conjecture is confirmed for both the paramagnetic and ferromagnetic phases numerically and analytically. In the ANNNI model, the conjecture could be analytically proven for a special case. Numerical methods are employed to investigate the conjecture's validity beyond this special case.  
\end{abstract}

\maketitle

\section{Introduction}
The quantum adiabatic theorem provides conditions under which a perturbed quantum system remains in its instantaneous eigenstate \cite{Born1928}. Beyond its fundamental conceptual significance, the theorem has been extensively used in condensed matter physics. It provided the basis for the introduction of the Berry phase \cite{Berry1984}, which plays a central role in the modern description of topological states of matter \cite{Budich2013, Avron1987}. Moreover, the adiabatic theorem is fundamental to several static and dynamical methods \cite{KingSmith1993, Albash2018} and has been used to establish numerous formal results \cite{Hastings2005, Henheik2021, Bachmann2017, Fabio2025}. Similarly, understanding the connection between ground states and eigenstates after a quantum quench can advance our understanding of non-equilibrium dynamics in quantum systems. Specifically, studying how ground states behave after quenches sheds light on energy distributions and phase transition dynamics. Hence, predicting overlaps between eigenstates can provide important insights into the mechanisms governing quench dynamics in non-equilibrium many-body systems. 

Quantum quenches and adiabatic time evolution represent two extremes in the study of quantum dynamics, both of which have been the subject of extensive theoretical and numerical research. Previous experimental \cite{Titum2019, Murayama2020, Bloch2008, Trotzky2012} and theoretical \cite{Rossini2020, Gautam2024, Bloch2008} works have explored the evolution of ground states in quenched systems and how memory of the system's initial state is retained. Quench dynamics have been used to study and identify phase transitions, either directly \cite{Haldar2021, Robertson2023} or through out-of-time-order correlators \cite{Heyl2018}. This work investigates the extent to which statements about ground state overlaps hold in the context of quantum quenches. In that sense, it seeks to extend the adiabatic theorem to maximally nonadiabatic changes.

After presenting the conjecture and its connection to the adiabatic theorem in Sec.~\ref{sec:conjecture} we present analytical results for two specific models, the transverse field Ising model (TFIM) and the axial next nearest neighbor Ising (ANNNI) model. For the TFIM we conduct a full analytical proof in Sec.~\ref{sec:TFIM}, whereas for the ANNNI model only a special case is proven analytically in Sec.~\ref{sec:ANNNI}. To investigate the conjecture's validity beyond this special case, section \ref{sec:numerical} contains a numerical analysis. In Sec.~\ref{sec:conclusion} we summarize our results and provide an outlook on future investigations.

\section{Conjecture}
\label{sec:conjecture}
The adiabatic theorem, formulated by Born and Fock in 1928, states \cite{Born1928},
\begin{quote}
    A physical system remains in its instantaneous eigenstate if a given perturbation is acting on it slowly enough and if there is a gap between the eigenvalue and the rest of the Hamiltonian's spectrum.
\end{quote}

This statement can be reexpressed in terms of quantum state overlaps: Consider a quantum system with an initial Hamiltonian $\mathcal{H}_i$ at time $t<0$, whose eigenstates are denoted by $|\psi_n\rangle$. Without loss of generality, assume the system is in its ground state, with an excitation gap $\delta E$ above it. At $t=0$, a perturbation acts on the system, such that the eigenstates evolve under the time-dependent Schrödinger equation to $|\psi_n(t>0)\rangle$. Let $\mathcal{H}_f$ denote the perturbed Hamiltonian, which typically has a different set of eigenstates $|\widetilde{\psi_n}\rangle$. For a perturbation with arbitrarily large ramping time $\tau$, the overlap of the time-evolved initial ground state and the final ground state approaches 1 arbitrarily closely:
\begin{align}
    \lim_{\tau \rightarrow \infty} |\langle\text{GS}(t>0) \mid\widetilde{\text{GS}} \rangle|^2 = 1 .
\end{align}
Any time evolution of a quantum system is called ``adiabatic'' if it happens \textit{slowly enough}. That is the case if the evolution is much slower than the inverse energy gap size, $\tau \gg (\delta E)^{-1}$.

The primary objective of this work is to investigate the relationship between eigenstates of different Hamiltonians when the system is quenched, i.e., when the adiabatic assumption is no longer valid. Explicitly, this paper aims to test the validity of a conjecture, which can be formulated as follows:
\begin{quote}
    The overlap of the initial ground state $|\text{GS}\rangle$ with any postquench eigenstate $|\widetilde{\psi_n}\rangle$ is maximal if $|\widetilde{\psi_n}\rangle$ is the postquench ground state, as long as both Hamiltonians are in the same phase:
    \begin{align}
    \label{eq:conjecture}
        \operatorname{max}_n |\langle\text{GS} \mid \widetilde{\psi_n}\rangle|^2 = |\langle\text{GS} \mid \widetilde{\text{GS}}\rangle|^2 .
    \end{align}    
\end{quote}
\begin{comment}
    REFEREE:
    Using the original formulation of the adiabatic theorem, one can think of a path in parameter space along which at any point the system is gapped. Then, one can change the Hamiltonian along this path slow enough such that the system remains in the instantaneous ground state. One way defining to be in the same phase would therefore be saying that there exists some path in the space of Hamiltonians between the pre- and postquench Hamiltonian, along which there is no gap-closing. Is this the notion of equal phases that the authors want to use for the conjecture?
\end{comment}
As a necessary condition, we restrict the validity of the conjecture to systems with a continuous spectrum. The notion of equal phases that we impose for this conjecture is the existence of a continuous path in the space of Hamiltonians with the same symmetries connecting the pre- and postquench Hamiltonians without any gap-closing. In other words, the system remains gapped at every point along this path. Since the concept of phase transitions is well-defined only in the continuous spectrum limit, the conjecture is expected to hold asymptotically for finite systems. The overlap in the conjecture above is closely related to the information-theoretic \textit{ground state fidelity} \cite{Wang2015,Sierant2019}
\begin{align}
\label{eq:fidelity}
    F(a, \delta a) &= |\left\langle \text{GS}(a) \mid \text{GS}(a+\delta a) \right\rangle|\\ \nonumber
    &= 1-\frac12 \chi_F(a)(\delta a)^2 + \mathcal{O}((\delta a)^3),
\end{align}
quantifying the resemblance of two ground states. Here, $a$ denotes a parameter of the Hamiltonian and $\delta a$ denotes a small change in $a$. For the thermodynamic limit one defines its intrinsic version as $\mathcal{F}(a, \delta a) = \lim_{N\rightarrow\infty} \frac{1}{N} \ln F(a, \delta a)$ \cite{Zhou2008,Zhou2008b,Gu2010} where $N$ denotes the system size.
An important quantity arising in the discussion of ground state fidelities is the \textit{fidelity susceptibility} $\chi_F(a)$, which is a measure of how rapidly the ground state changes under perturbations of a control parameter. If the fidelity susceptibility per lattice site is an analytic function, a finite parameter region in which the ground state overlap remains the largest must exist. Away from quantum critical points and (quantum) phase transitions the fidelity susceptibility is expected to remain an analytic function of the control parameter. It is widely accepted in the community––and supported by much numerical evidence––that the fidelity susceptibility can diverge only in the vicinity of phase transitions \cite{Zanardi2006, Venuti2007} and quantum critical points \cite{Mukherjee2011, Thakurathi2012}. However, there is no mathematically rigorous proof for it. Because even this weaker statement about the fidelity susceptibility has not been proven (yet), one should not expect to establish a proof for the more general conjecture formulated here. Currently, the only possible approach to verify or disprove the conjecture is with numerical studies, just as is the case for the fidelity susceptibility. Nevertheless, we believe that our conjecture holds in a finite parameter region sufficiently far from phase transitions and quantum critical points.

\section{Transverse Field Ising Model}
\label{sec:TFIM}
The Hamiltonian of the one-dimensional TFIM is given by
\begin{align}
\label{eq:TFIM}
    H_{\mathrm{TFIM}} = -J \sum_j \left(\sigma_j^z \sigma_{j+1}^z + h\sigma_j^x \right),
\end{align}
where $\sigma^x$ and $\sigma^z$ denote Pauli matrices. The parameter $J$ denotes an overall energy scale, subsequently set equal to 1 for convenience, while $h>0$ defines the strength of the transverse magnetic field. At zero temperature, this model undergoes a phase transition from a gapped ferromagnet to a gapped paramagnet at a critical value of $h=J$ \cite{Silva2008}.

After Jordan--Wigner \cite{JordanWigner1993} and Bogoliubov \cite{Bogoliubov1958} transformations, the Hamiltonian in momentum space reads
\begin{align}
    H_{\mathrm{TFIM}} = \sum_{k>0} \big[ \epsilon_k \big(\gamma_k^{\dagger} \gamma_k - \gamma_{-k}\gamma_{-k}^{\dagger}\big) - h \big],
\end{align}
where $\epsilon_k = \sqrt{h^2-2 h \cos (k)+1}$ and $\gamma_k^{\dagger}$ and $\gamma_k$ denote the Bogoliubov operators
\begin{align}
    \gamma_k &= \cos(\theta_k) c_k + i \sin(\theta_k) c_{-k}^{\dagger},\\
    \gamma_k^{\dagger} &= \cos(\theta_k) c_k^{\dagger} - i \sin(\theta_k) c_{-k},
\end{align}
which are in turn defined via the Fourier-transformed operators of a rotated Jordan--Wigner transformation
\begin{align}
    \sigma_i^x &= 1-2 c_i^{\dagger} c_i, \\
    \sigma_i^z &= -\prod_{j<i}\big(1-2 c_j^{\dagger} c_j\big)\big(c_i+c_i^{\dagger}\big),
\end{align}
using the Fourier convention
\begin{align}
    c_k = \frac{1}{\sqrt{L}} \sum_j e^{-i k j} c_j, \quad c_k^{\dagger} = \frac{1}{\sqrt{L}} \sum_j e^{i k j} c_j^{\dagger}.
\end{align}

This is the starting point for the subsequent calculations, where the general idea is to quench from an initial (\textit{prequench}) Hamiltonian $H_i$ to a final (\textit{postquench}) Hamiltonian $H_f$:
\begin{align}
    H_i \ \xrightarrow{t\ll (\delta E)^{-1}}\ H_f .
\end{align}
In general, these Hamiltonians $H_i$ and $H_f$ have different eigenbases and the prequench ground state  expressed in the postquench eigenbasis takes the form
\begin{align}
\label{eq:GS}
    |\text{GS}(h)\rangle = \frac{1}{\mathcal{N}} \exp{\bigg(\sum_{k>0} B^*(k) \gamma_k^{\dagger}\gamma_{-k}^{\dagger}\bigg)}|0,0\rangle.
\end{align}
The coefficient $B(k)$ is defined by \cite{Silva2008}
\begin{align}
\label{eq:Bk}
    B(k) := -i \frac{V_k}{U_k} =-i \tan[\theta_k(h_i)-\theta_k(h_f)],
\end{align}
where the coefficients $V_k$ and $U_k$ are given by
\begin{flalign}
    &U_k = \cos[\theta_k(h_i)]\cos[\theta_k(h_f)] + \sin[\theta_k(h_i)]\sin[\theta_k(h_f)],\nonumber&\\
    &V_k = \cos[\theta_k(h_i)]\sin[\theta_k(h_f)] - \sin[\theta_k(h_i)]\cos[\theta_k(h_f)].&
\end{flalign}
These expressions and the ground state \eqref{eq:GS} are valid for any value of $h$, and therefore apply in both the paramagnetic and ferromagnetic phase. The angles $\theta_k(h)$ are given by the parametrization of the Bogoliubov transformation \cite{Silva2008}
\begin{flalign}
\label{eq:theta_h}
    &\tan(2\theta_k) = \frac{\sin k}{h - \cos k} \ \Leftrightarrow \ \theta_k(h) = \frac{1}{2} \arctan\left(\frac{\sin k}{h - \cos k}\right)&
\end{flalign}
and are limited to the interval $[0,\frac{\pi}{2}]$. As $k = 2\pi\frac{n}{L}$, with $n \in \mathbb{N}$ in the interval $[1,\frac{L}{2}]$ and $L$ being the spin chain length, the values for $k$ cover half a period from 0 to $\pi$.

For the excited states of the postquench Hamiltonian in its own eigenbasis, there are three different possibilities:
\begin{align}
    |0,1\rangle &= \gamma_k^{\dagger} |0,0\rangle \label{eq:ex+},\\
    |1,0\rangle &= \gamma_{-k}^{\dagger} |0,0\rangle \label{eq:ex-},\\
    |1,1\rangle &= \gamma_k^{\dagger}\gamma_{-k}^{\dagger} |0,0\rangle.
\label{eq:exex}
\end{align}
The excitations in Eqs.~\eqref{eq:ex+} and \eqref{eq:ex-} involve unpaired creation operators, resulting in zero overlap with the prequench ground state. The more interesting excitations are those in Eq.~\eqref{eq:exex}, as they involve paired creation operators and can have non-zero overlaps with the prequench ground state.

We first calculate the ground state overlap:
\begin{align}
\label{eq:GS00}
    |\langle \text{GS}(h)\mid 0,0\rangle|^2 &= \bigg|\frac{1}{\mathcal{N}} \langle 0,0\mid \exp{\left(\sum_{k>0}B(k) \gamma_{-k}\gamma_k\right)} \mid 0,0 \rangle\bigg|^2\nonumber\\
    & = \frac{1}{\mathcal{N}^2} \bigg|\langle 0,0\mid \prod_{k>0} \bigg(1+ B(k) \gamma_{-k}\gamma_k\bigg) \mid 0,0 \rangle\bigg|^2\nonumber\\
    & = \frac{1}{\mathcal{N}^2}.
\end{align}
Then, we compute the overlap of the prequench ground state with the postquench excited state as defined in Eq.~\eqref{eq:exex}:
\begin{small}
\begin{align}
    \Big|\langle \text{GS}&(h)\mid 1,1\rangle\Big|^2\nonumber\\
    &= \frac{1}{\mathcal{N}^2} \left|\langle 0,0\mid \exp{\left(\sum_{k>0}B(k) \gamma_{-k}\gamma_k\right)} \gamma_{-k'}^{\dagger}\gamma_{k'}^{\dagger} \mid 0,0 \rangle\right|^2\nonumber\\
    & = \frac{1}{\mathcal{N}^2} \Big|B(k')\Big|^2.
\end{align}
\end{small}
Inserting the definition \eqref{eq:Bk} for $B(k)$ into this expression yields
\begin{align}
\label{eq:GS11}
    \bigg|\langle \text{GS}(h)\mid 1,1\rangle\bigg|^2 = \frac{1}{\mathcal{N}^2} \tan^2\left[\theta_k(h_i) - \theta_k(h_f)\right].
\end{align}
Expression \eqref{eq:GS11} must be smaller than the ground state overlap \eqref{eq:GS00} for the conjecture to be correct; hence
\begin{align}
    \Big|\langle\text{GS}(h)\mid 1,1\rangle\Big|^2 \overset{!}{<} \Big|\langle\text{GS}(h)\mid 0,0\rangle\Big|^2\\
    \Leftrightarrow \frac{1}{\mathcal{N}^2} \tan^2\left[\theta_k(h_i) - \theta_k(h_f)\right] \overset{!}{<} \frac{1}{\mathcal{N}^2}\\
    \Leftrightarrow \tan^2\left[\theta_k(h_i) - \theta_k(h_f)\right] \overset{!}{<} 1.
\label{eq:condition}
\end{align}
Condition \eqref{eq:condition} is fulfilled if and only if the argument of $\tan^2(x)$ lies in the interval $[-\frac{\pi}{4}, \frac{\pi}{4}]$. We conclude that
\begin{align}
    \big|\theta_k(h_i) - \theta_k(h_f)\big| \overset{!}{<} \frac{\pi}{4}.
\end{align}
Taking the parametrization of $\theta_k(h)$ that was given in Eq.~\eqref{eq:theta_h} and using the inverse tangent function
\begin{align}
    \tan^{-1}(2\theta_k) = \begin{cases}
        \arctan(2\theta_k) , \qquad \ \, \ 2\theta_k>0,\\
        \arctan(2\theta_k)+\pi , \quad 2\theta_k<0,
    \end{cases}
\end{align}
imply, for the definition of $\theta_k(h)$,
\begin{flalign}
\label{eq:theta_cases}
    &\theta_k(h) = \begin{cases}
        \frac{1}{2}\arctan(\frac{\sin k}{h - \cos k}), \qquad \ \ \, \frac{\sin k}{h - \cos k} > 0,\\
        \frac{1}{2}[\arctan(\frac{\sin k}{h - \cos k}) + \pi], \ \, \frac{\sin k}{h - \cos k} < 0.&
    \end{cases}
\end{flalign}
The condition on the left-hand side of Eq.~\eqref{eq:condition} depends on three variables, $h_i$, $h_f$, and $k$. Its maximum with respect to each of these variables is computed, leading to the distinction of three different scenarios:
\begin{enumerate}
    \item[(1)] Both Hamiltonians lie in the paramagnetic phase. Without loss of generality, it is assumed that ${1<h_i<h_f}$. In the paramagnetic phase, ${\theta_k = \frac{1}{2}\arctan(\frac{\sin k}{h - \cos k})\ \forall\ h_i, h_f, k}$ because ${\frac{\sin k}{h - \cos k} > 0}$ for all values of ${h_i, h_f \in (1,\infty)}$ and ${k \in [0,\pi]}$.
    
    Here $\arctan(\frac{\sin k}{h_i - \cos k})$ is a monotonically decreasing function in $h$ for all $k \in [0,\pi]$. Therefore, the maximal difference of $\theta_k(h_i)$ and $\theta_k(h_f)$ is reached for maximally different $h_i$ and $h_f$. We choose $h_i$ to be $1+\epsilon$ with $\epsilon \rightarrow 0$ and $h_f$ to be $x\rightarrow\infty$ and obtain
    \begin{align}
        &\max_{h_i,h_f,k}\left(\left|\theta_k(h_i) - \theta_k(h_f)\right|\right) = \frac{\pi}{4}.
    \end{align}
    As the function approaches the limit from below, conjecture \eqref{eq:conjecture} is fulfilled for all possible parameter combinations of $h_i$, $h_f$ and $k$ in the paramagnetic phase.
    
    \item[(2)] Both Hamiltonians lie in the ferromagnetic phase. Without loss of generality, it is assumed that ${0 \le h_i < h_f < 1}$. For ${h \in [0,1)}$, we rewrite Eq.~\eqref{eq:theta_cases} as
    \begin{align}
        \theta_k(h) = \begin{cases}
        \frac{1}{2}\arctan(\frac{\sin k}{h - \cos k}), \qquad \quad \ \ k > \arccos(h),\\
        \frac{1}{2}\big[\arctan(\frac{\sin k}{h - \cos k}) + \pi\big], \quad k < \arccos(h).
    \end{cases}
    \end{align}
    We consider ${\arccos(h) < k = \arccos(h) + \nu}$ with $\nu > 0$:
    \begin{align}
        &\quad \max_{h_i,h_f,k}\left[\left|\theta_k(h_i) - \theta_k(h_f)\right|\right]\\
        &= \max_{h_i,h_f,\nu}\bigg[\frac{1}{2}\bigg|\arctan\left(\frac{\sin[\arccos(h_i) + \nu]}{h_i - \cos[\arccos(h_i) + \nu]}\right) \nonumber\\
        &\quad \quad \quad \ - \arctan\left(\frac{\sin[\arccos(h_f) + \nu]}{h_f - \cos[\arccos(h_f) + \nu]}\right)\bigg|\bigg].\nonumber
    \end{align}
    As $\arctan\left(\cdot\right)$ is a monotone function in $h$ for all $\nu \in (0,\pi]$, the maximal difference of $\theta_k(h_i)$ and $\theta_k(h_f)$ is reached for maximally different $h_i$ and $h_f$. Thus, we choose $h_i$ to be zero and $h_f$ to be $1-\epsilon$ with $\epsilon \rightarrow 0$, such that
    \begin{align}
        &\max_{h_i,h_f,k}\left[\left|\theta_k(h_i) - \theta_k(h_f)\right|\right]\nonumber\\
        &=\max_{\nu}\bigg[\frac{1}{2}\bigg|\arctan\left(\frac{\sin(\frac{\pi}{2} + \nu)}{\cos(\frac{\pi}{2} + \nu)}\right)\nonumber\\
        &\quad \quad \quad + \arctan\left(\frac{\sin(\nu)}{1 - \cos(\nu)}\right)\bigg|\bigg]
    \end{align}
    follows. For $k \in [0, \pi]$ to be fulfilled, it becomes apparent that $\nu \overset{!}{\in} (0,\frac{\pi}{2}]$ and
    \begin{align}
        \max_{h_i,h_f,k}\left[\left|\theta_k(h_i) - \theta_k(h_f)\right|\right] = \frac{\pi}{8} < \frac{\pi}{4}.
    \end{align}
    Hence, the conjecture \eqref{eq:conjecture} is confirmed for this choice of $\theta_k(h)$.
    
    \item[(3)] Again, both Hamiltonians lie in the ferromagnetic phase. As before, without loss of generality, it is assumed that $0 < h_i < h_f < 1$. However, now we consider the case of $\arccos(h) > k = \arccos(h) - \nu$ with $\nu > 0$. From $k \in [0,\pi]$ it follows that $\nu \overset{!}{\in} [-\frac{\pi}{2},0)$.
    \begin{small}
    \begin{align}
        &\max_{h_i,h_f,k}\left[\left|\theta_k(h_i) - \theta_k(h_f)\right|\right]\\
        &=\max_{h_i,h_f,\nu}\bigg[\frac{1}{2}\bigg|\arctan\left(\frac{\sin[\arccos(h_i) - \nu]}{h_i - \cos[\arccos(h_i) - \nu]}\right)+\pi \nonumber\\
        &\quad \quad \quad \quad - \arctan\left(\frac{\sin[\arccos(h_f) - \nu]}{h_f - \cos[\arccos(h_f) - \nu]}\right)+\pi\bigg|\bigg].\nonumber
    \end{align}
    \end{small}
    As $\arctan(\cdot)$ is a monotone function in $h$ for all $\nu \in [-\frac{\pi}{2},0)$, the maximal difference of $\theta_k(h_i)$ and $\theta_k(h_f)$ is again reached for maximally different $h_i$ and $h_f$. Thus we again choose $h_i$ to be zero and $h_f$ to be $1-\epsilon$ with $\epsilon \rightarrow 0$, such that
    \begin{align}
        \max_{h_i,h_f,k}\left[\left|\theta_k(h_i) - \theta_k(h_f)\right|\right] = \frac{\pi}{8} < \frac{\pi}{4},
    \end{align}
    and conjecture \eqref{eq:conjecture} is confirmed for this second choice of $\theta_k(h)$.
\end{enumerate}
One can easily check that condition \eqref{eq:condition} is not necessarily satisfied once the phase transition is crossed. In fact, in the vicinity of the critical point at $h=1$, the condition is highly sensitive to even small changes in $h$. For instance, choosing ${h_i < h_f}$ (without loss of generality) in different phases, with ${h_i = 1-\epsilon}$ and ${h_f = 1+\epsilon}$, an $\epsilon$ of order $10^{-2}$ is already sufficient to violate condition \eqref{eq:condition} for some $k$. As $\epsilon$ increases further, the set of momenta for which the condition fails grows correspondingly.

Of course, not only first-order but also higher-order excitations can be considered. In general, these are of the form 
\begin{align}
    |\zeta,\zeta\rangle &= \gamma_k^{\dagger}\gamma_{k'}^{\dagger} ... \gamma_{-k}^{\dagger}\gamma_{-k'}^{\dagger} ... |0,0\rangle.
\end{align}
However, they have an even smaller overlap with the prequench ground state than the excitation \eqref{eq:exex}, as each excitation pair contributes another factor further diminishing the overlap with the prequench ground state.

As shown above, we could analytically confirm the conjecture for the TFIM.

\section{Axial Next Nearest Neighbor Ising Model}
\label{sec:ANNNI}
\begin{figure}
    \centering
    \includegraphics[width=0.45\textwidth]{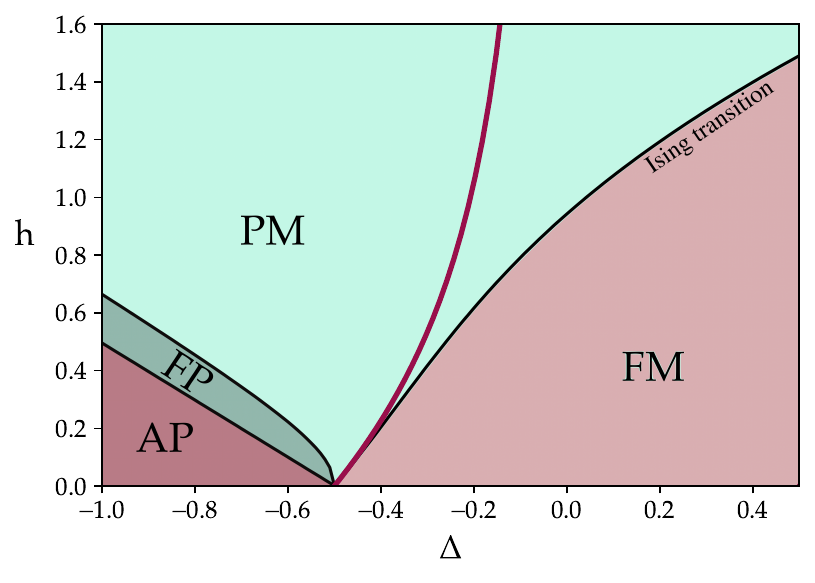}
    \caption{Phase diagram of the ANNNI model. The red line within the paramagnetic phase indicates the Peschel--Emery line. Adapted from \cite{Karrasch2013}.}
    \label{fig:ANNNI_phasediag}
\end{figure}
The second model investigated in this work is a modification of the transverse field Ising model, incorporating additional next nearest neighbor interactions. It is known as the anisotropic or axial next nearest neighbor Ising model and has been studied extensively in the last few decades \cite{Elliott1961, Fisher1980, Karrasch2013, Haldar2021, Robertson2023}. Alongside the nearest neighbor coupling $J$ and the coupling $h$ to an external magnetic field, the interaction between next nearest neighbors can be tuned by a parameter $\Delta$ in the Hamiltonian \cite{Karrasch2013}
\begin{figure}
    \centering
    \begin{subfigure}{0.5\textwidth}
        \centering
        \begin{overpic}[width=0.88\textwidth]{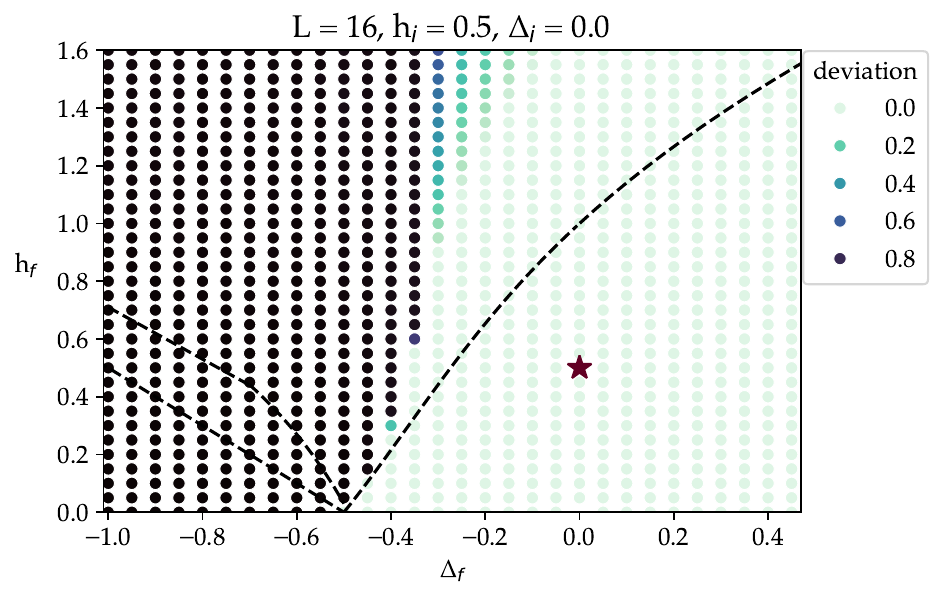}
        \put(-6,60){\textbf{(a)}}
        \end{overpic}
    \end{subfigure}

    \begin{subfigure}{0.5\textwidth}
        \centering
        \begin{overpic}[width=0.88\textwidth]{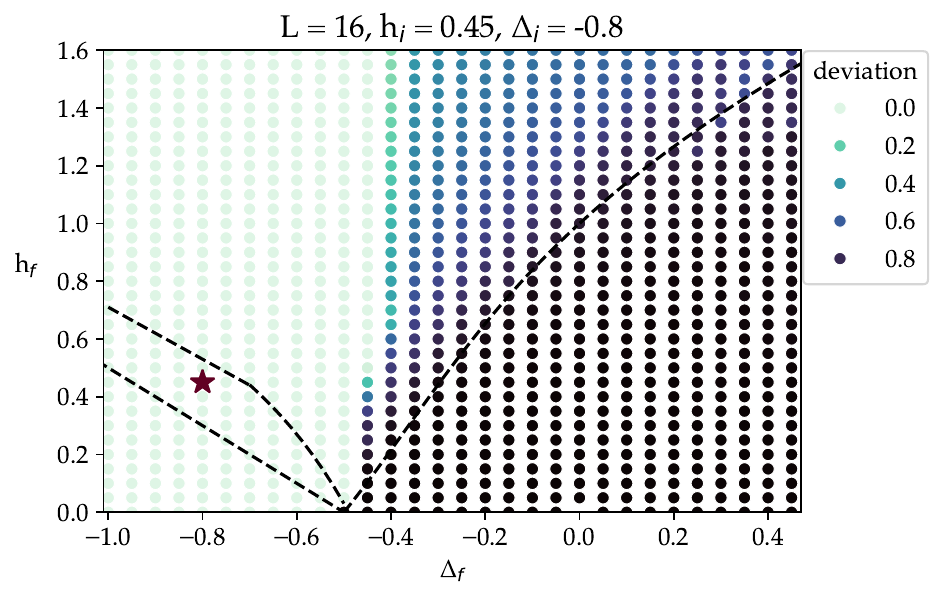}
        \put(-6,60){\textbf{(b)}}
        \end{overpic}
    \end{subfigure}

    \begin{subfigure}{0.5\textwidth}
        \centering
        \begin{overpic}[width=0.88\textwidth]{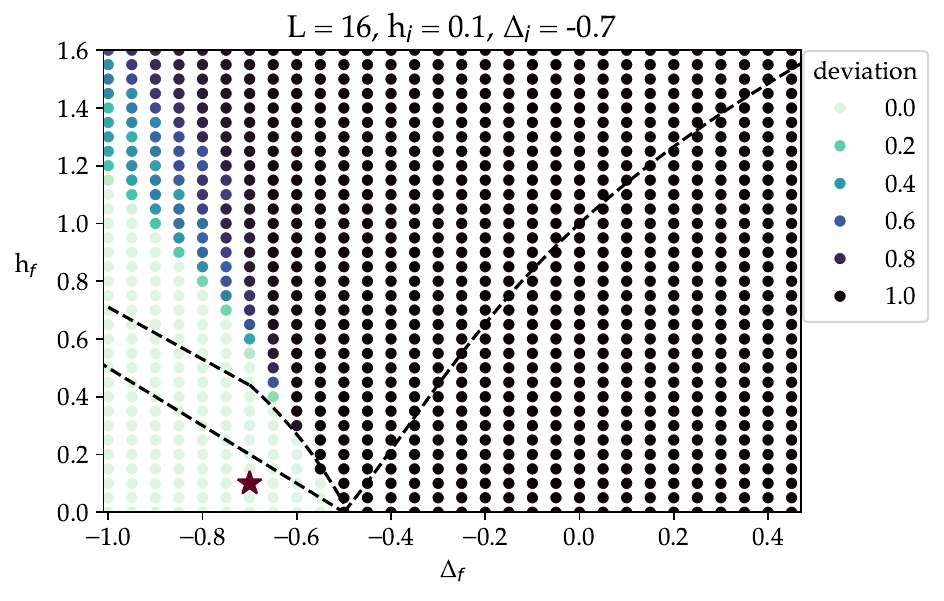}
        \put(-6,60){\textbf{(c)}}
        \end{overpic}
    \end{subfigure}
    \caption{Quenches within the (a) ferromagnetic and (b) floating phases and (c) the antiphase.}
    \label{fig:ANNNI-FM}
\end{figure}
\begin{align}
\label{eq:Ham_ANNNI}
    H_{\mathrm{ANNNI}} = -J \sum_{j}\left(\sigma_{j}^{z} \sigma_{j+1}^{z}+\Delta \sigma_{j}^{z} \sigma_{j+2}^{z}+h \sigma_{j}^{x}\right).
\end{align}
This interaction strength $\Delta$ is also known as the ``frustration'' parameter \cite{Cea2024}. The ANNNI model is the simplest model that incorporates quantum fluctuations (coupling $h$) and frustrated exchange interactions (coupling $\Delta$). As such, it serves as a fundamental model for exploring the dynamics between magnetic ordering, frustration, and disordering effects \cite{Cea2024}.\\
The competition between interactions gives rise to a more complex phase diagram, accommodating four different phases, as illustrated in Fig. \ref{fig:ANNNI_phasediag}. These phases are the paramagnetic (PM) phase, the ferromagnetic (FM) phase, a critical incommensurate floating phase (FP), and an antiphase (AP).\\
In contrast to the TFIM, the ANNNI model is not integrable \cite{Karrasch2013}, except along two lines in the phase diagram: the trivial line with $\Delta = 0$ (which corresponds to the TFIM) and the so-called Peschel--Emery (PE) line \cite{Peschel1981}. The PE line is a disorder line along which the Hamiltonian factorizes into local Hamiltonians $H_j^{\operatorname{loc}}$, such that it takes the form \cite{Katsura2015}
\begin{align}
\label{eq:ANNNI_PE}
    H_{\mathrm{ANNNI}}^{\mathrm{PE}} = \sum_j H_j^{\operatorname{loc}} .
\end{align}
The ground state of this factorized Hamiltonian is exactly degenerate and can be expressed as a product state \cite{Mahyaeh2018}. The model described by Eq.~\eqref{eq:ANNNI_PE} is frustration-free, as the ground state of the full Hamiltonian $H_{\mathrm{ANNNI}}^{\mathrm{PE}}$ minimizes each local Hamiltonian $H_j^{\operatorname{loc}}$ independently \cite{Katsura2015}. In the ANNNI model, the PE line is described by \cite{Peschel1981}
\begin{align}
    h = h^{*} := \Delta - \frac{1}{4\Delta}.
\end{align}
\begin{figure}
    \centering
    \begin{tabular}{cc}
        % First row
        \begin{overpic}[width=0.45\linewidth, trim=0.15cm 0 2.32cm 0, clip]{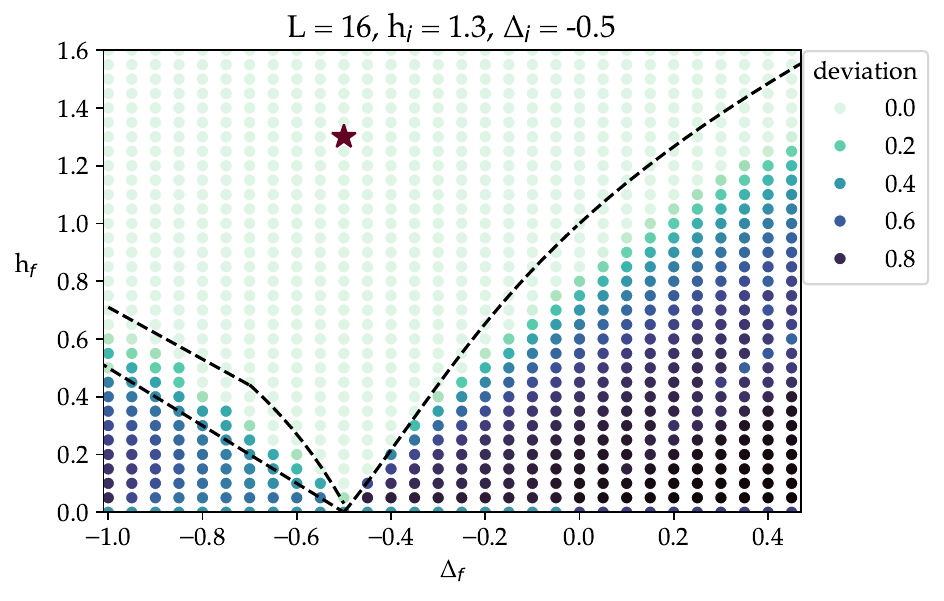}
        \put(-9,65){\textbf{(a)}}
        \end{overpic} &
        \begin{overpic}[width=0.45\linewidth, trim=0 0 2.32cm 0, clip]{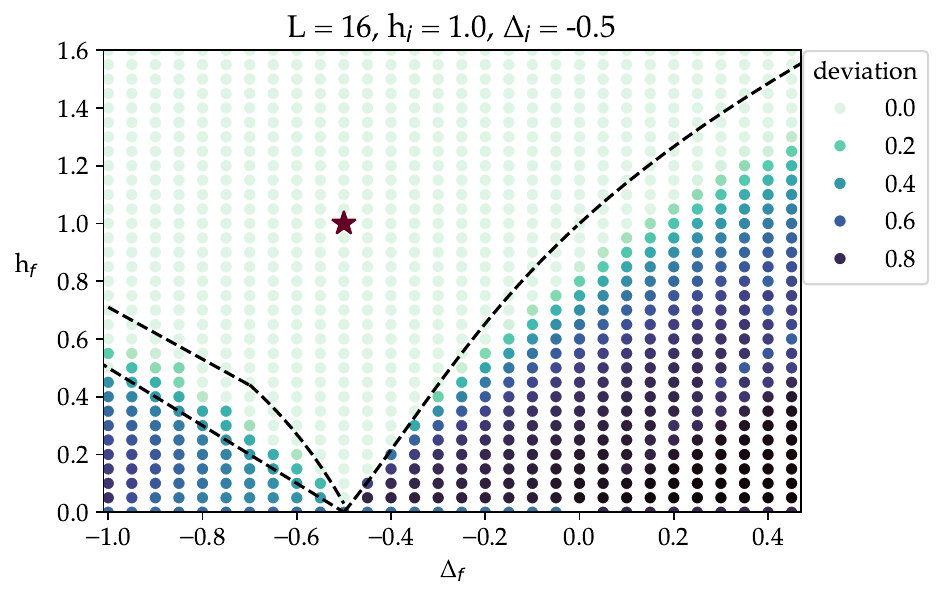}
        \put(-6,65){\textbf{(b)}}
        \end{overpic} \\
        % Second row
        \begin{overpic}[width=0.45\linewidth, trim=0.15cm 0 2.32cm 0, clip]{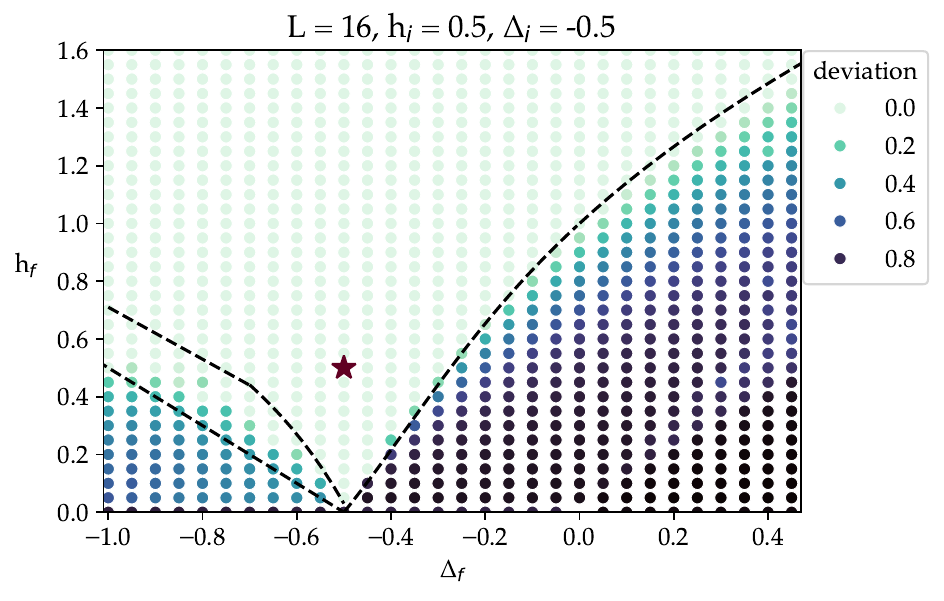}
        \put(-9,65){\textbf{(c)}}
        \end{overpic} & 
        \begin{overpic}[width=0.45\linewidth, trim=0 0 2.32cm 0, clip]{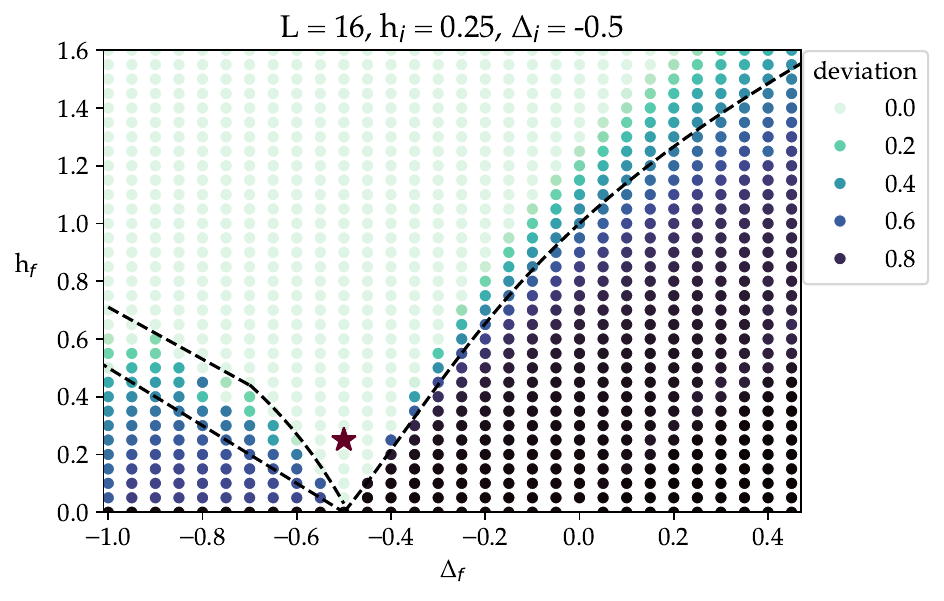}
        \put(-6,65){\textbf{(d)}}
        \end{overpic}
        % \captionsetup{labelfont={color=white}}
        % \caption{}
        \label{fig:13d}
    \end{tabular}
    \caption{Quenches within the paramagnetic phase of the starting point successively approaching the quadruple point.}
    \label{fig:PM-approach-quad}
\end{figure}
The Hamiltonian of the ANNNI model can be mapped onto an interacting quantum Ising model by dual mapping and a subsequent rotation \cite{Mahyaeh2018}. After the further application of a Jordan--Wigner transformation, this Hamiltonian is equivalent to the Kitaev--Hubbard chain \cite{Mahyaeh2018, Katsura2015, Sela2011}:
\begin{align}
\label{eq:Ham-ANNNI}
    H_\mathrm{ANNNI} &= J \sum_{j=1}^{L}\big[ h(c_{j}^{\dagger} -c_{j})(c_{j+1}^{\dagger}+c_{j+1})\\
    &+ \Delta (1-2 c_{j}^{\dagger} c_{j})(1-2 c_{j+1}^{\dagger} c_{j+1}) - (1-2 c_{j}^{\dagger} c_{j})\big].\nonumber
\end{align}
Due to the quartic interaction arising from the next-nearest neighbor coupling in the spin-$\frac{1}{2}$ system, the ANNNI model does not have a simple analytical solution like the TFIM.

As the Hamiltonian factorizes, the ground state along the PE line is a direct product state \cite{Katsura2015}:
\begin{align}
\label{eq:gs-factorized}
    |\text{GS}_i^{(\pm)}\rangle = \tens_{j=1}^{L} \big(\left|\uparrow\right\rangle_j \pm \alpha \left|\downarrow\right\rangle_j \big),
\end{align}
where ${\alpha = \sqrt{\cot{\left(\frac{\theta^*}{2}\right)}}}$ with ${\theta^* = \arctan{\left(h^*\right)}} = \arctan{\left(\Delta - \frac{1}{4\Delta}\right)}$ \cite{Katsura2015}. This ground state could be normalized by the factor $(1 + \alpha^2)^{-L/2}$, but we omit normalization because it is not required for the following discussion. We see that $\alpha \ge 1$  along the entire frustration-free line.
\begin{figure}
    \centering
    \begin{subfigure}{0.5\textwidth}
        \centering
        \begin{overpic}[width=0.75\textwidth, trim=0 0 2.32cm 0, clip]{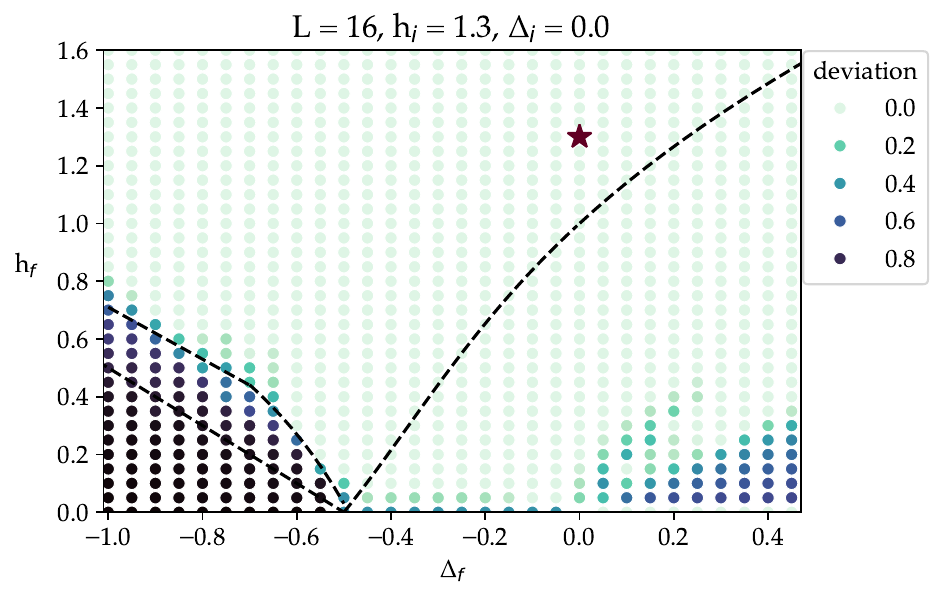}
        \put(-6,67){\textbf{(a)}}
        \end{overpic}
        % \captionsetup{labelfont={color=white}}
        % \caption{}
        \label{fig:14a}
    \end{subfigure}

    \begin{subfigure}{0.5\textwidth}
        \centering
        \begin{overpic}[width=0.75\textwidth, trim=0 0 2.32cm 0, clip]{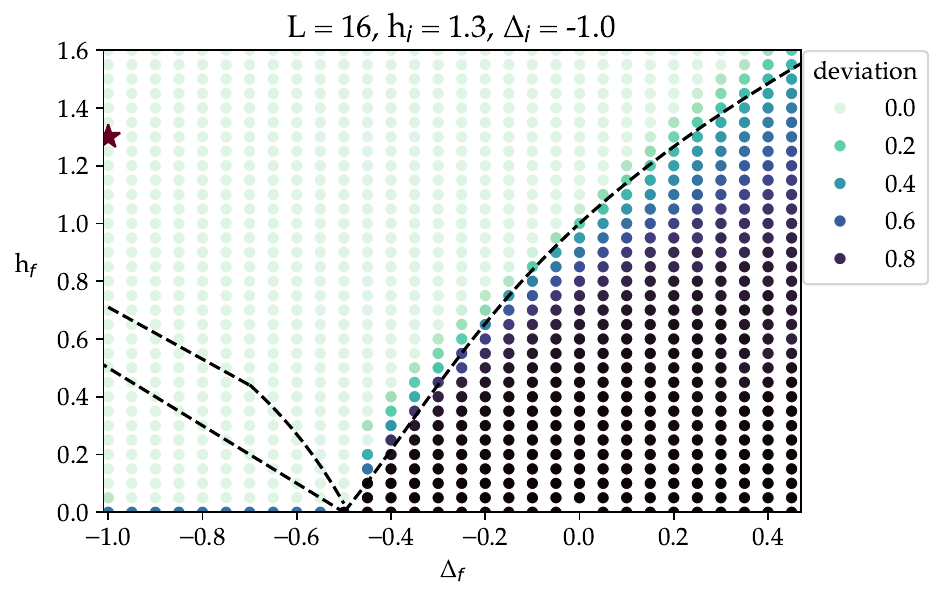}
        \put(-6,67){\textbf{(b)}}
        \end{overpic}
    \end{subfigure}

    \begin{subfigure}{0.5\textwidth}
        \centering
        \begin{overpic}[width=0.75\textwidth, trim=0 0 2.32cm 0, clip]{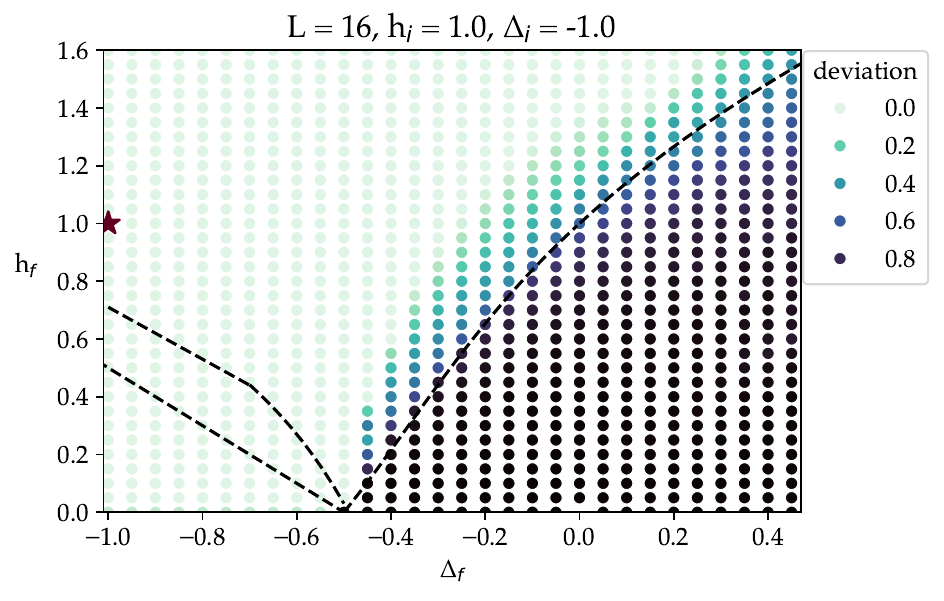}
        \put(-6,67){\textbf{(c)}}
        \end{overpic}
        % \captionsetup{labelfont={color=white}}
        % \caption{}
        \label{subfig:fail}
    \end{subfigure}
    \caption{Other quenches within the paramagnetic phase showing violations of the conjecture.}
    \label{fig:mother_fail}
\end{figure}

\begin{figure}
    \centering
    \begin{subfigure}{0.5\textwidth}
        \centering
        \begin{overpic}[width=0.75\textwidth, trim=0 0 2.32cm 0, clip]{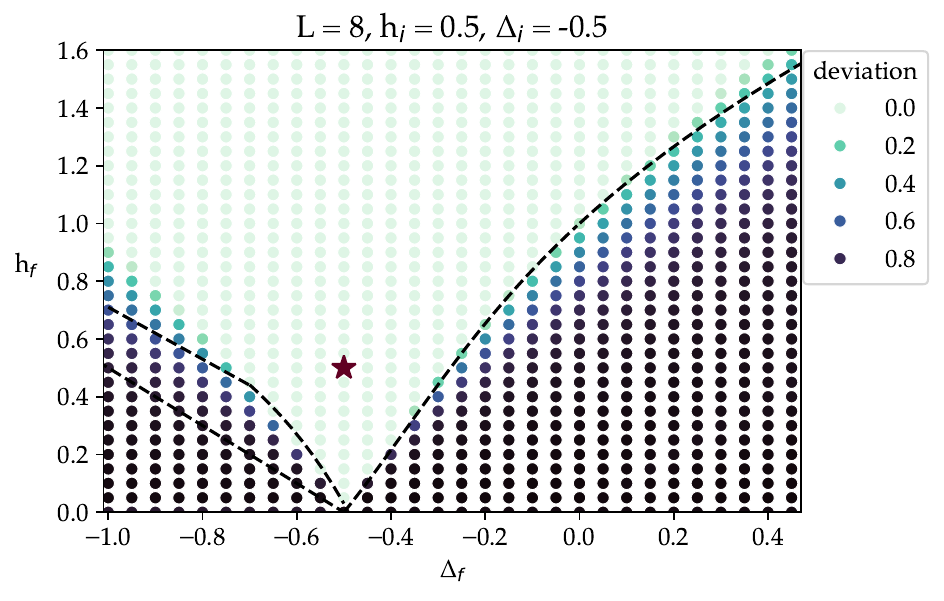}
        \put(-6,67){\textbf{(a)}}
        \end{overpic}
        % \captionsetup{labelfont={color=white}}
        % \caption{}
        \label{fig:17a}
    \end{subfigure}

    \begin{subfigure}{0.5\textwidth}
        \centering
        \begin{overpic}[width=0.75\textwidth, trim=0 0 2.32cm 0, clip]{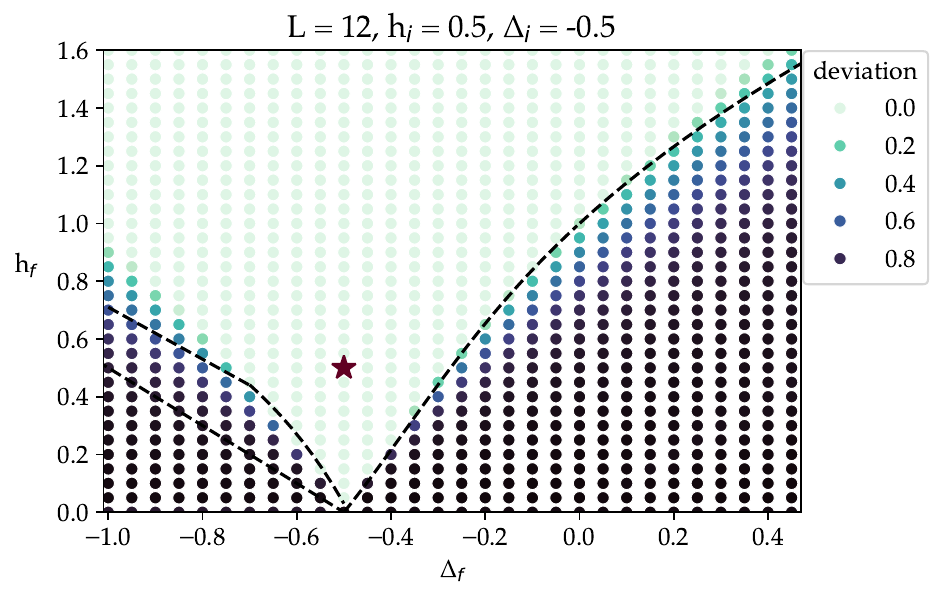}
        \put(-6,67){\textbf{(b)}}
        \end{overpic}
        % \captionsetup{labelfont={color=white}}
        % \caption{}
        \label{fig:17b}
    \end{subfigure}

    \begin{subfigure}{0.5\textwidth}
        \centering
        \begin{overpic}[width=0.75\textwidth, trim=0 0 2.32cm 0, clip]{L16_Delta-0.5_h0.5.pdf}
        \put(-6,67){\textbf{(c)}}
        \end{overpic}
    \end{subfigure}
    \caption{Quenches starting in the paramagnetic phase at $(\Delta,h) = (-0.5,0.5)$ for different system sizes.}
    \label{fig:violations-size-dep}
\end{figure}
Excitations of the ground state \eqref{eq:gs-factorized} can be created by replacing one or more of the local ground states with their orthogonal complements, ${(\left|\uparrow\right\rangle \pm \alpha \left|\downarrow\right\rangle)^{\perp} = (\pm \alpha\left|\uparrow\right\rangle - \left|\downarrow\right\rangle)}$
\begin{align}
    \hspace{-0.2cm}|\psi^{(\pm)}\rangle = \tens_{j=1}^N \left( |\uparrow \rangle_j \pm \alpha |\downarrow\rangle_j \right) \tens_{l=N+1}^L \left( \pm \alpha |\uparrow \rangle_l -|\downarrow\rangle_l \right)
\end{align}
for $N<L$. These states, like the ground state, are not normalized. Note, that these states are exact eigenstates only along the PE line.

The overlap between two ground states along the PE line with positions characterized by $\alpha_i$ and $\alpha_f$, respectively, then reads
\begin{align}
\nonumber
    \big|\langle \text{GS}_i^{(\pm)} &| \text{GS}_f^{(\pm)} \rangle\big|^2\\
    =&\ \bigg| \tens_{j=1}^L \left(\left\langle\uparrow\right|_j \pm \alpha_i \left\langle\downarrow\right|_j \right) \tens_{l=1}^L \left(\left|\uparrow\right\rangle_l \pm \alpha_f \left|\downarrow\right\rangle_l \right) \bigg|^2\nonumber\\
    =&\ \big| (1+\alpha_i\alpha_f)^L \big|^2 = (1+\alpha_i\alpha_f)^{2L},
\end{align}
and the overlap between a ground state and an excited state yields
\begin{align}
\nonumber
    \big|\langle \text{GS}_i^{(\pm)} | \psi_f^{(\pm)} \rangle\big|^2 = \bigg| &\tens_{j=1}^L \left(\left\langle\uparrow\right|_j \pm \alpha \left\langle\downarrow\right|_j \right)\tens_{l=1}^N \left( |\uparrow \rangle_l \pm \alpha |\downarrow\rangle_l \right)\nonumber\\
    &\hspace{-0.15cm} \times \tens_{m=N+1}^L \left( \pm \alpha |\uparrow \rangle_m -|\downarrow\rangle_m \right) \bigg|^2\\
    =&\ (1 + \alpha_i\alpha_f)^{2N} (\pm\alpha_f \mp \alpha_i)^{2(L-N)}.\nonumber
\end{align}
For the conjecture to hold, the ratio of these two overlaps should be smaller than one:
\begin{equation}
\begin{aligned}
\hspace{-0.15cm}
\label{eq:final-condition}
    \big|\langle\text{GS}_i |\text{GS}_f \rangle\big|^2 \overset{!}{>} \big|\langle\text{GS}_i |\psi_f \rangle\big|^2 \Leftrightarrow \Bigg|\frac{\langle\text{GS}_i |\psi_f \rangle}{\langle\text{GS}_i |\text{GS}_f \rangle}\Bigg|^2 \overset{!}{<} 1.
\end{aligned}
\end{equation}
\begin{figure}
    \centering
    \begin{subfigure}{0.5\textwidth}
        \centering
        \begin{overpic}[width=0.9\textwidth]{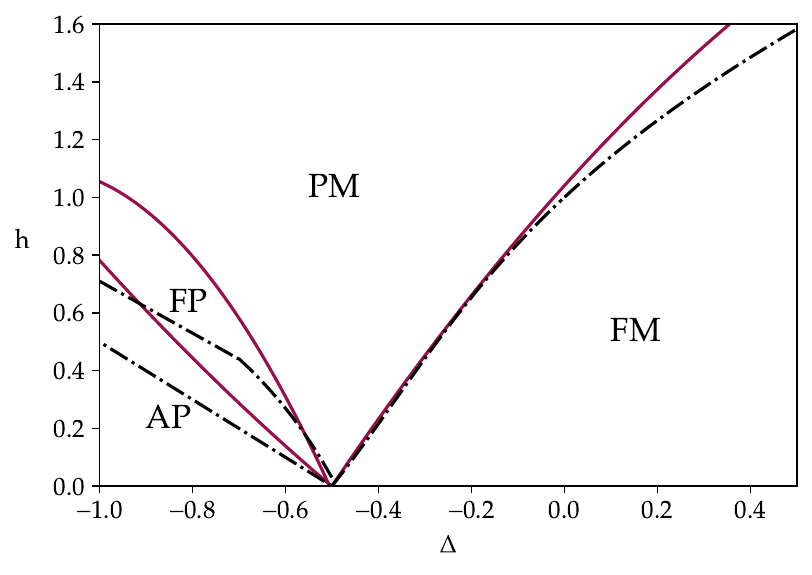}
        \put(-6,67){\textbf{(a)}}
        \end{overpic}
    \end{subfigure}

    \begin{subfigure}{0.5\textwidth}
        \centering
        \begin{overpic}[width=0.9\textwidth]{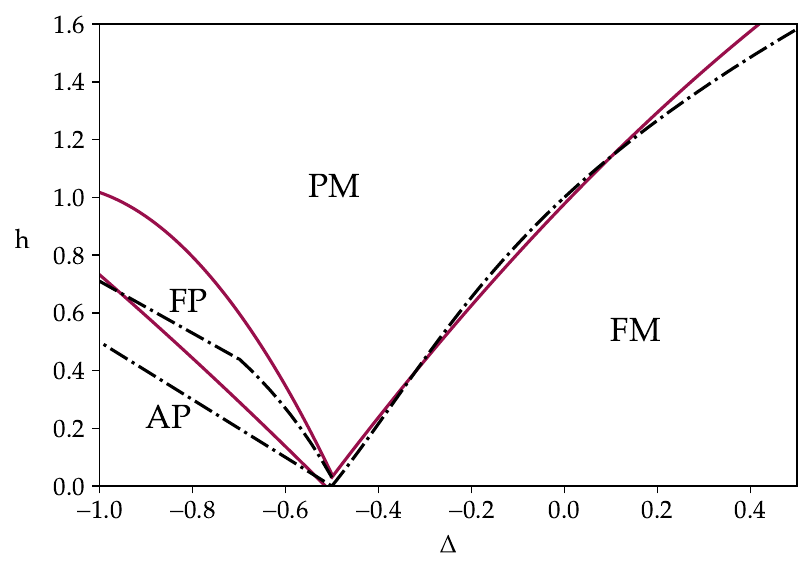}
        \put(-6,67){\textbf{(b)}}
        \end{overpic}
    \end{subfigure}
    \caption{Finite-size shifted phase boundaries of the ANNNI model when keeping the parameters (a) $h$ and (b) $\Delta$ fixed, respectively. The dashed lines mark the DMRG-calculated phase boundaries of the infinite-sized system \cite{Karrasch2013}, while the red lines show the shifted transition lines.}
    \label{fig:shifted-phase-diags}
\end{figure}

\begin{figure}
    \centering
    \begin{subfigure}{0.5\textwidth}
        \begin{overpic}[width=0.75\textwidth]{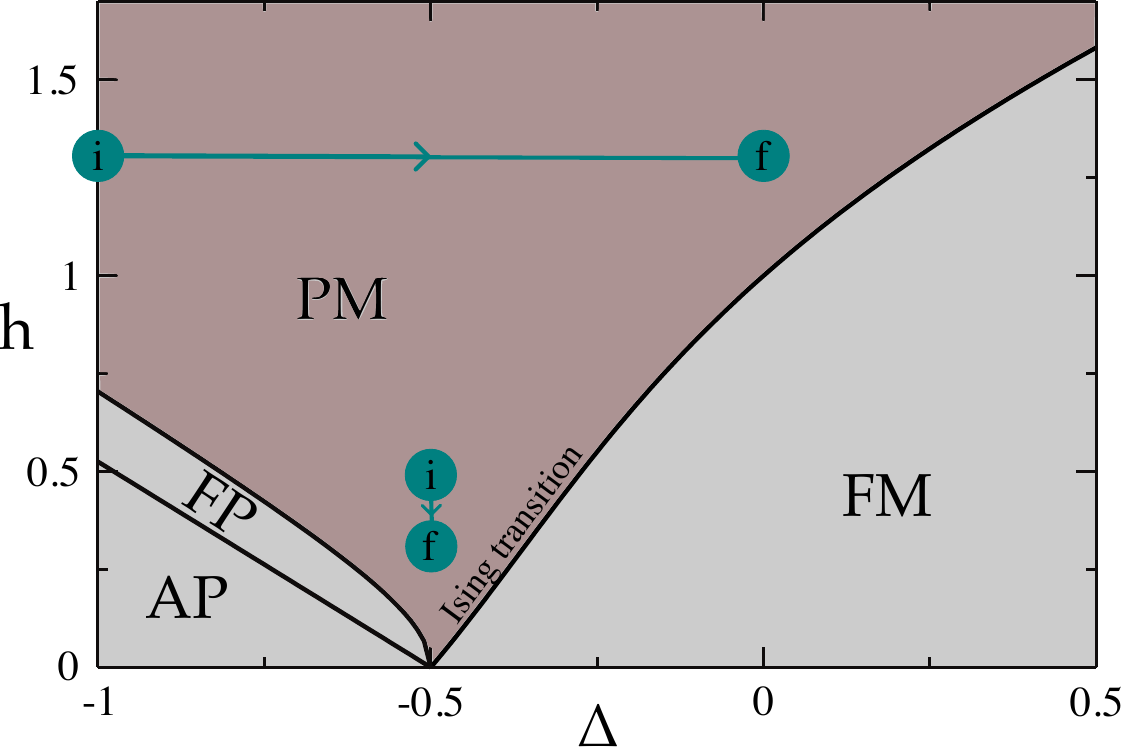}
        \put(-9,63){\textbf{(a)}}
        \end{overpic}
        % \captionsetup{labelfont={color=white}}
        % \caption{}
        \label{subfig:PM-quenches}
    \end{subfigure}
    \begin{subfigure}{0.5\textwidth}
        \centering
        \begin{overpic}[width=0.95\textwidth]{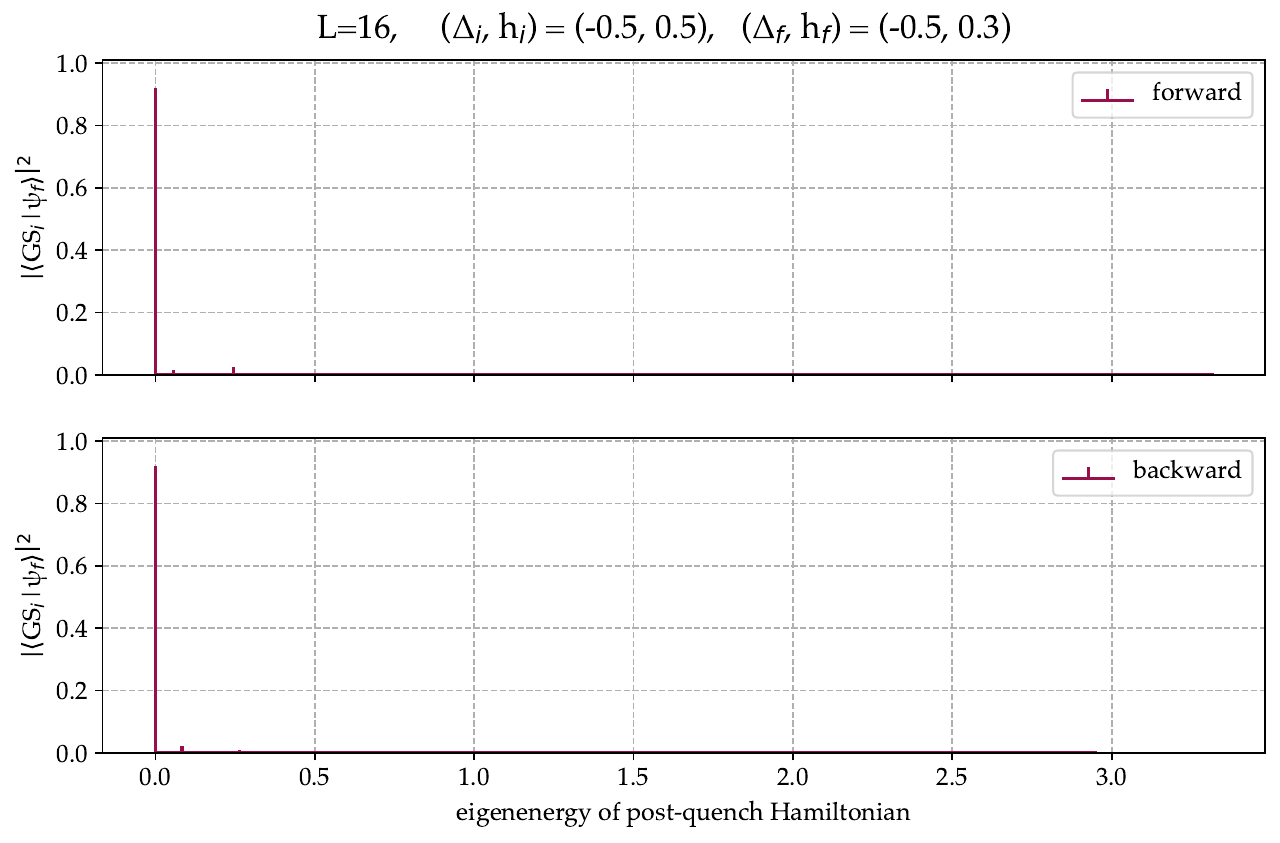}
        \put(-5,60){\textbf{(b)}}
        \end{overpic}
    \end{subfigure}

    \begin{subfigure}{0.5\textwidth}
        \centering
        \begin{overpic}[width=0.95\textwidth]{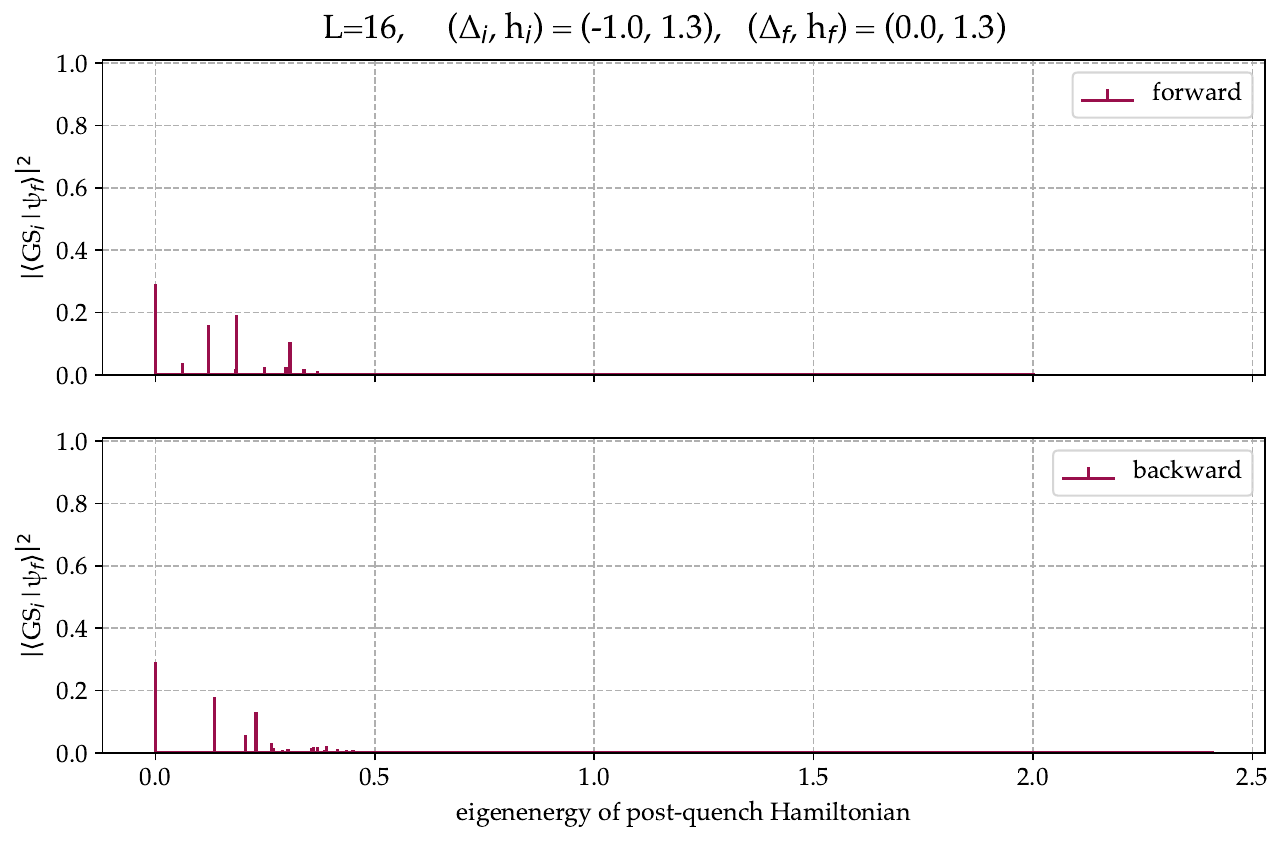}
        \put(-5,60){\textbf{(c)}}
        \end{overpic}
    \end{subfigure}
    \caption{Quenches within the paramagnetic phase. (a) Phase diagram indicating the starting $i$ and ending $f$ points of the quenches. (b), (c) Magnitude of the overlap between the prequench ground state at $i$ and each eigenstate of the finite system at $f$, shown for the vertical (b) and horizontal (c) quenches marked in (a). A quench from $i$ to $f$ is referred to as ``forward.'' Adapted from Karrasch and Schuricht \cite{Karrasch2013}.}
    \label{fig:PM}
\end{figure}
Calculating this ratio gives
\begin{align}
\nonumber
    \bigg|\frac{\langle\text{GS}_i |\psi_f \rangle}{\langle\text{GS}_i |\text{GS}_f \rangle}\bigg|^2 &= \frac{(1\pm\alpha_i\alpha_f)^{2N} (\pm \alpha_f \mp \alpha_i)^{2(L-N)}}{(1+\alpha_i\alpha_f)^{2L}}\\
    &= \bigg(\frac{\pm \alpha_f \mp \alpha_i}{1+\alpha_i\alpha_f}\bigg)^{2(L-N)}.
\label{eq:final-result-ANNNI}
\end{align}
To decide whether this expression is larger or smaller than 1, we recall that $\alpha \in [1,\infty)$. Thus, we can distinguish four cases.
\begin{enumerate}[itemsep=-3pt]
    \item[(1)] $\alpha_i \rightarrow 1$, \text{ and } $\alpha_f \rightarrow \infty$.
    \item[(2)] $\alpha_i \rightarrow \infty$, \text{ and } $\alpha_f \rightarrow 1$.
    \item[(3)] $\alpha_i \rightarrow 1$, \text{ and } $\alpha_f \rightarrow 1$.
    \item[(4)] $\alpha_i \rightarrow \infty$, \text{ and } $\alpha_f \rightarrow \infty$.
\end{enumerate}
The limit for all four cases tends to zero and is thus smaller than 1. Therefore, the conjecture that can be rephrased into condition \eqref{eq:final-condition} is fulfilled.

The two product states $|\text{GS}^{(+)} \rangle$ and $|\text{GS}^{(-)} \rangle$ are both exact ground states and are exactly degenerate. Further, they are related by the global $\mathbb{Z}_2$ spin-flip symmetry but they belong to different symmetry sectors. Thus, their respective excitations $|\psi^{(+)} \rangle$ and $|\psi^{(-)} \rangle$ also belong to different sectors. As a consequence, the physically relevant overlaps are those within the same sector (i.e., between states carrying the same superscript).

We want to emphasize that for calculating the overlap, only the number of flipped spins is relevant; the number of domain walls does not influence the result. Eigenstates that differ from an ``ordered'' excited state (e.g., $\left|\uparrow\uparrow\uparrow\uparrow\downarrow\downarrow\right\rangle$) may have more domain walls and, consequently, a different excitation energy. However, this does not affect the quantum mechanical overlap.

Finally, Eq.~\eqref{eq:final-result-ANNNI} confirms that condition \eqref{eq:final-condition} is satisfied along the entire frustration-free line. Thus, the conjecture can be verified analytically for the special case of a quench along this line.
\begin{figure}
    \centering
    \begin{subfigure}{0.5\textwidth}
        \centering
        \begin{overpic}[width=0.75\textwidth]{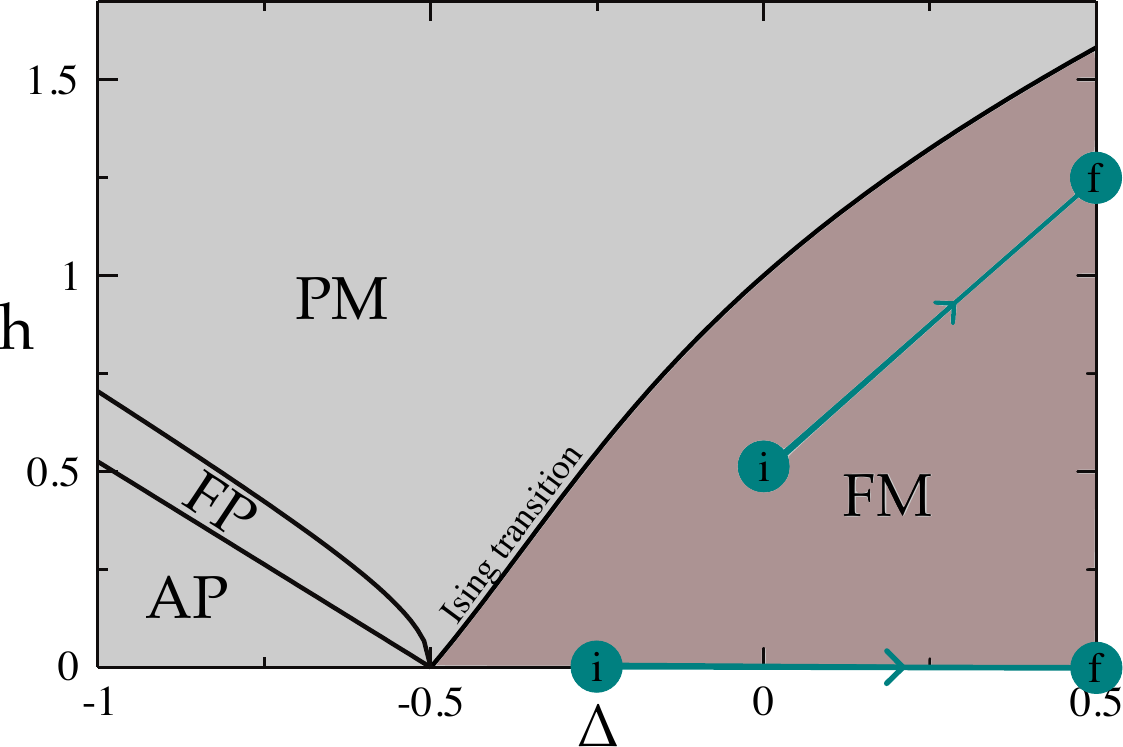}
        \put(-9,63){\textbf{(a)}}
        \end{overpic}
    \end{subfigure}

    \begin{subfigure}{0.5\textwidth}
        \centering
        \begin{overpic}[width=0.95\textwidth]{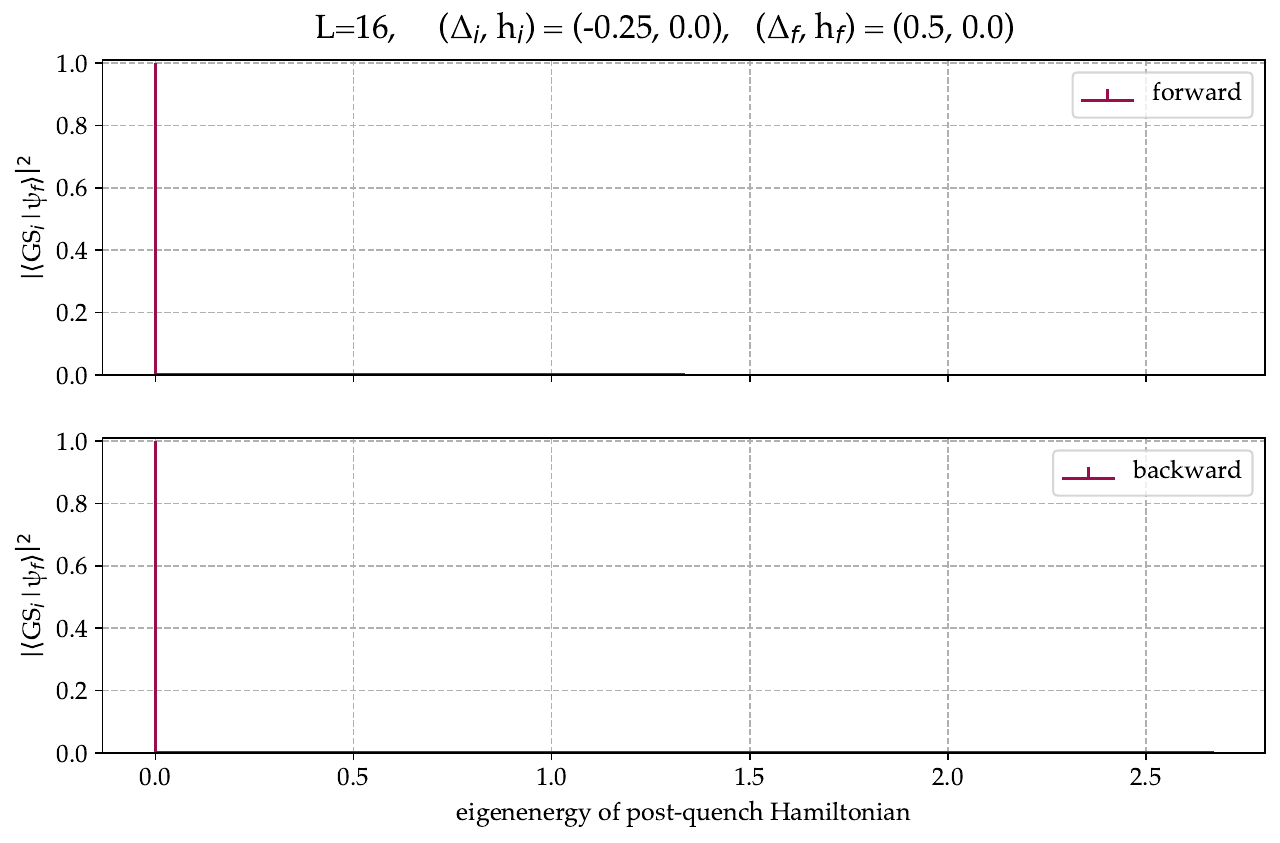}
        \put(-5,63){\textbf{(b)}}
        \end{overpic}
    \end{subfigure}

    \begin{subfigure}{0.5\textwidth}
        \centering
        \begin{overpic}[width=0.95\textwidth]{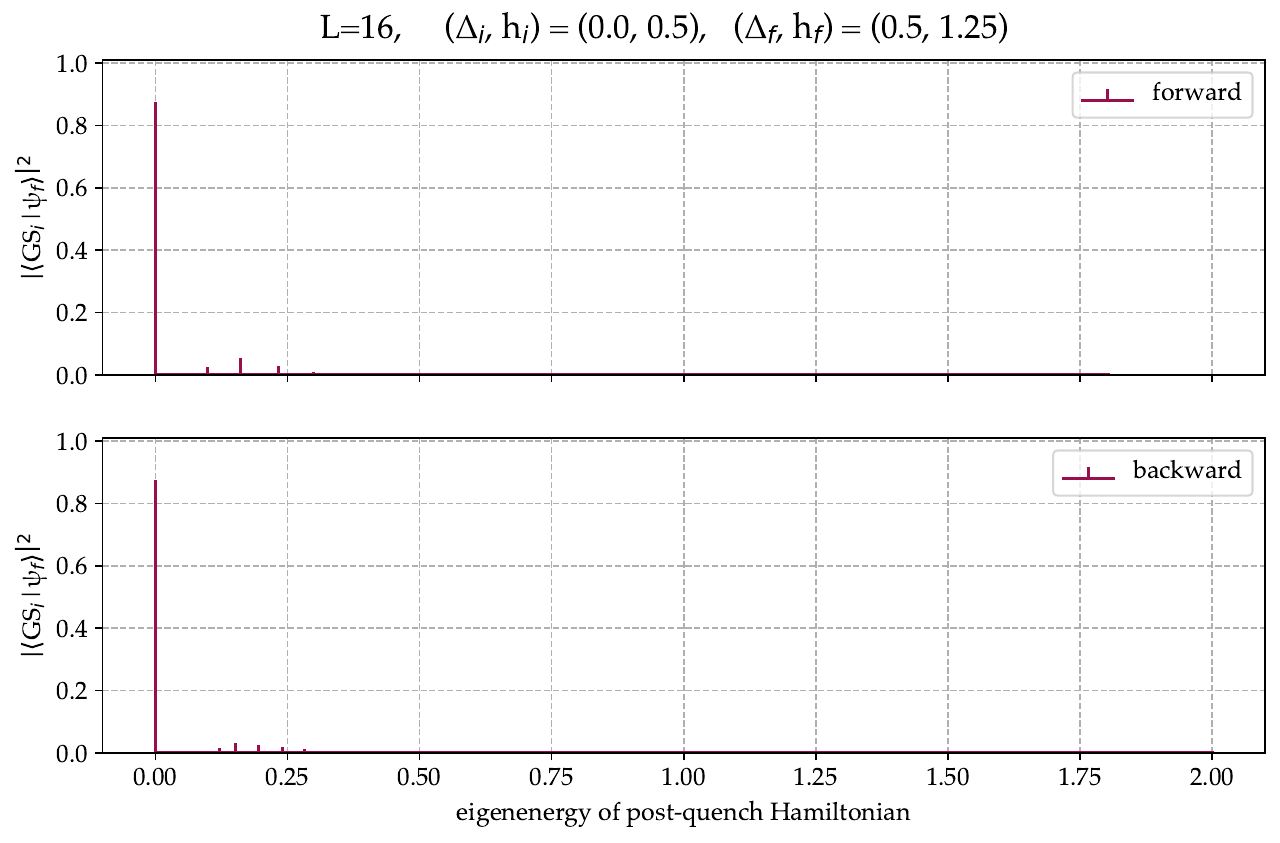}
        \put(-5,63){\textbf{(c)}}
        \end{overpic}
    \end{subfigure}
    \caption{Quenches within the ferromagnetic phase. The phase diagram in (a) shows the starting and ending points of the quenches. The quench from $i$ to $f$ is referred to as ``forward.'' Adapted from Karrasch and Schuricht \cite{Karrasch2013}.}
    \label{fig:FM}
\end{figure}

\begin{figure}
    \centering
    \begin{subfigure}{0.5\textwidth}
        \centering
        \begin{overpic}[width=0.75\textwidth]{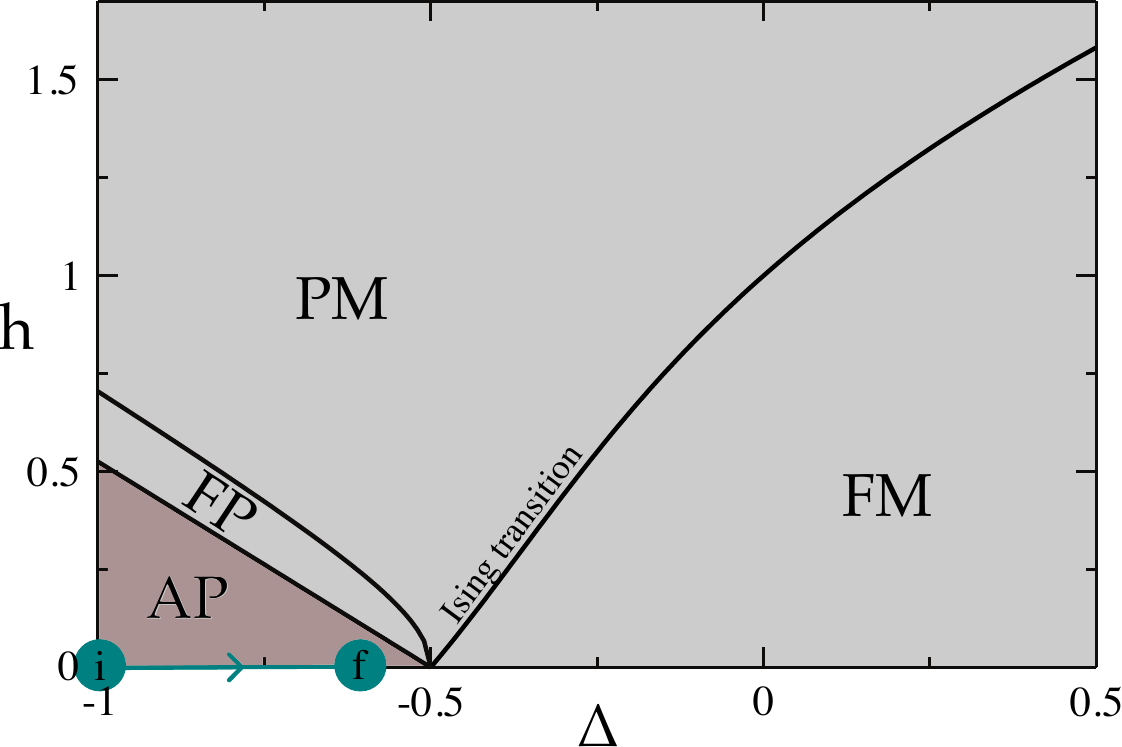}
        \put(-9,63){\textbf{(a)}}
        \end{overpic}
    \end{subfigure}

    \begin{subfigure}{0.5\textwidth}
        \centering
        \begin{overpic}[width=0.9\textwidth]{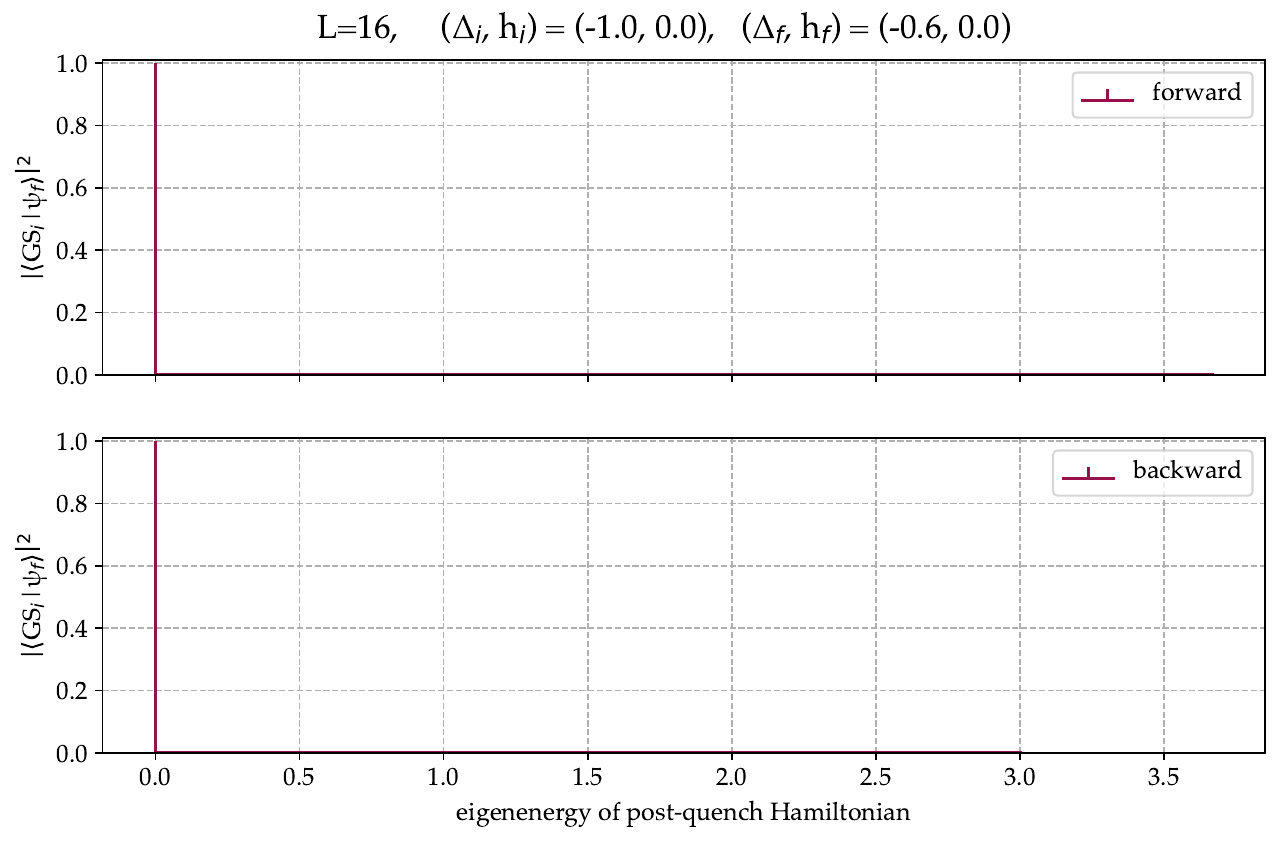}
        \put(-5,63){\textbf{(b)}}
        \end{overpic}
    \end{subfigure}
    \caption{Quench within the antiphase. The phase diagram in (a) shows the starting and ending points of the quench. The quench from $i$ to $f$ is referred to as ``forward.'' Adapted from Karrasch and Schuricht \cite{Karrasch2013}.}
    \label{fig:AP}
\end{figure}

\section{Numerical Analysis of the ANNNI model}
\label{sec:numerical}
To investigate the conjecture's validity beyond this special case, we conducted a numerical analysis of the ANNNI model. We present results obtained through exact diagonalisation (ED). Given the constraints of the available computational resources, the largest accessible system size for these computations is $L=16$ spins. The presentation of the results is structured according to the phase diagram shown in Fig.~\ref{fig:ANNNI_phasediag}. Each point in the phase diagram corresponds to a parameter combination that uniquely defines the postquench Hamiltonian. For clarity and ease of interpretation, dashed lines are used to reconstruct the phase boundaries. The quench starting points are indicated above each plot and marked by a star in the phase diagram. The results are displayed using a color code based on the normalized \textit{deviation} of the ground state overlap from the maximal overlap, defined as
\begin{align}
    \text{deviation} = \frac{|\text{GS}\rangle\ \text{overlap} - \text{max overlap}}{\text{max. overlap}}.
\end{align}
In this visualization, a very bright teal color indicates that the ground state overlap equals the maximal overlap, while darker colors signify a greater deviation between these two quantities.
\begin{figure}
    \centering
    \begin{subfigure}{0.5\textwidth}
        \centering
        \begin{overpic}[width=0.75\textwidth]{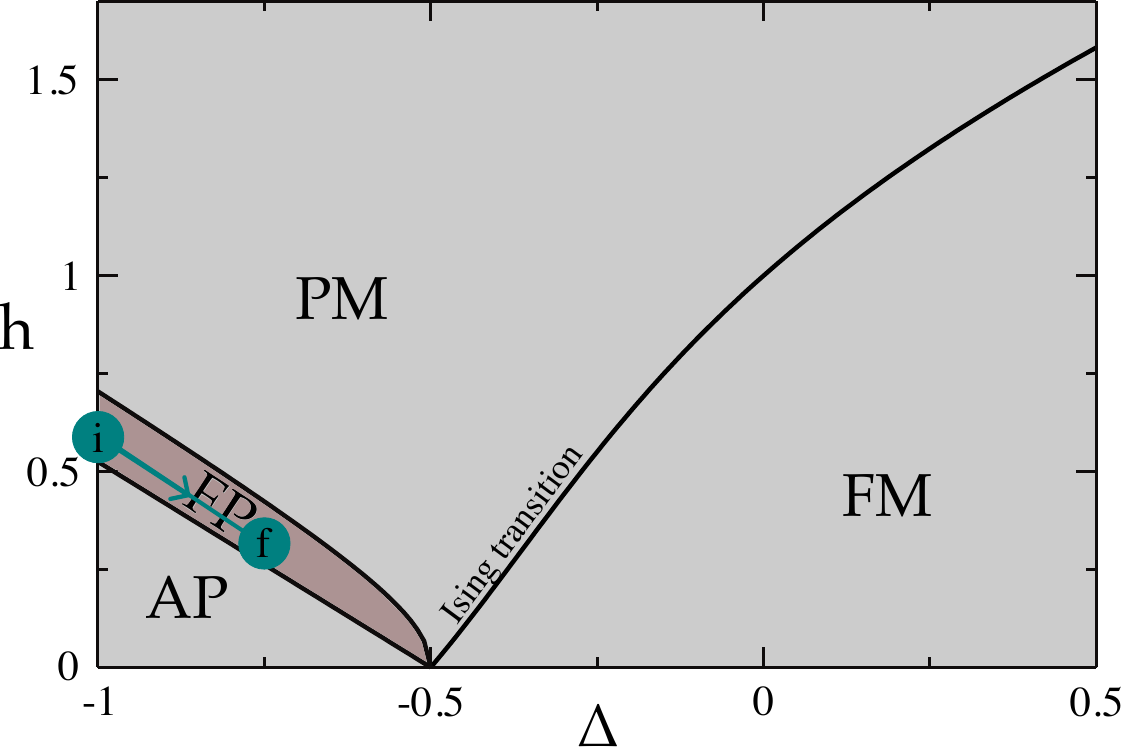}
        \put(-9,63){\textbf{(a)}}
        \end{overpic}
    \end{subfigure}

    \begin{subfigure}{0.5\textwidth}
        \centering
        \begin{overpic}[width=0.9\textwidth]{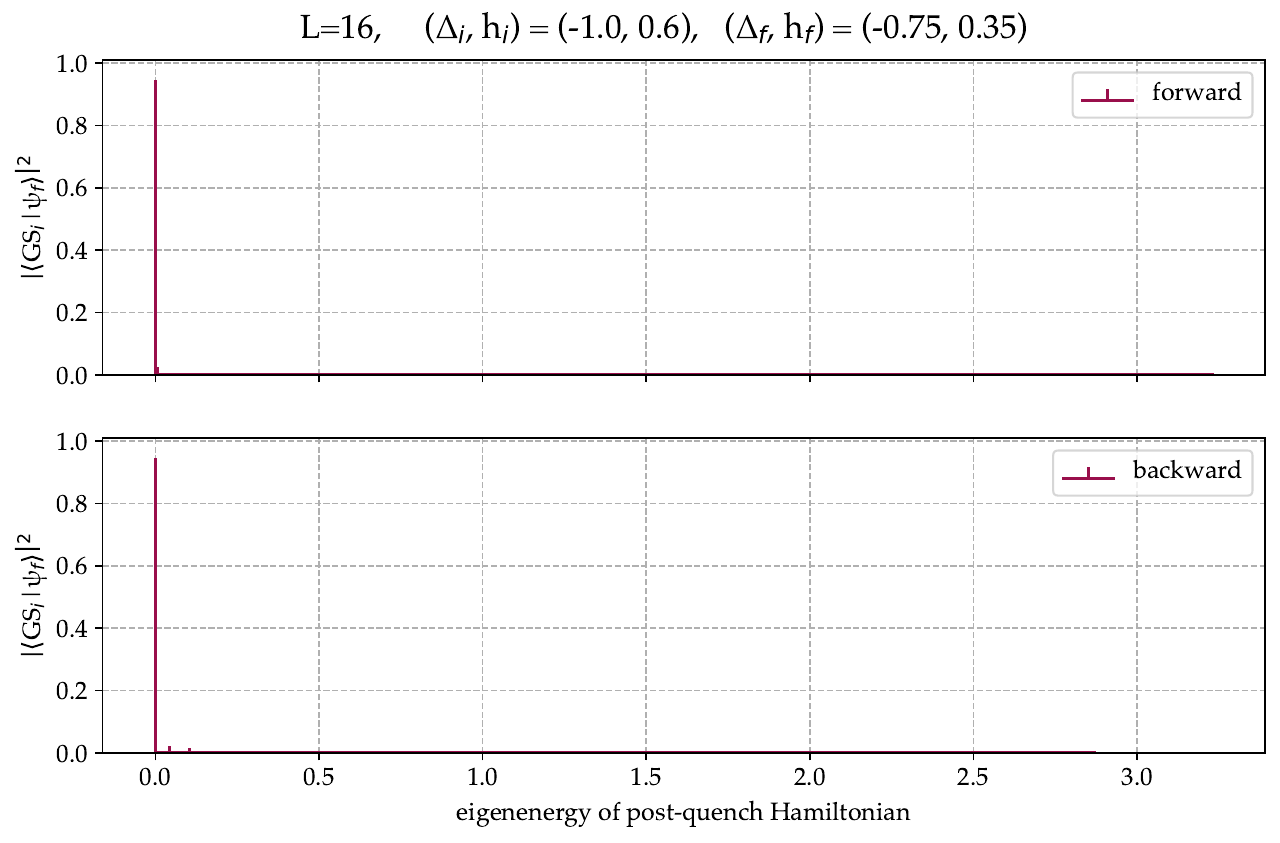}
        \put(-5,63){\textbf{(b)}}
        \end{overpic}
    \end{subfigure}
    \caption{Quench within the floating phase. The phase diagram in (a) shows the starting and ending points of the quench. The quench from $i$ to $f$ is referred to as ``forward.'' Adapted from Karrasch and Schuricht \cite{Karrasch2013}.}
    \label{fig:FP}
\end{figure}

Within the phase where the quench starts, all points should ideally be displayed in the brightest color for the conjecture to hold true. Any darker color within that phase indicates a violation of the conjecture, as it signifies that the maximal overlap is not provided by the ground state overlap in such cases.

For all tested quench configurations within the ferromagnetic and floating phases and antiphase, the numerical data consistently support the conjecture (Fig.~\ref{fig:ANNNI-FM}). However, for prequench ground states in the paramagnetic phase, the numerical results are partly inconclusive.

Figure \ref{fig:PM-approach-quad} shows a starting point at ${\Delta_i = -0.5}$ successively approaching the quadruple point. When the quench starts far from this multiphase point, the numerical results support the conjecture. However, as the quench starting point moves closer to the quadruple point, violations of the conjecture begin to appear close to the Ising phase transition (Fig.~\ref{fig:PM-approach-quad}). Figure \ref{fig:violations-size-dep} explores whether these violations could be caused by finite-size effects, as a shifted position of the phase boundaries for smaller systems is a known phenomenon in the ANNNI model \cite{Cea2024}.

The regions where the conjecture is violated become progressively smaller as the system size increases, suggesting that these violations might be due to finite-size effects. A more detailed investigation of this surmise can be conducted by computing the finite-size shifted locations of the phase transitions.

The concept of a ``phase transition'' is not well-defined in finite systems, particularly when the considered systems are very small. Strictly speaking, the term applies to only systems in the thermodynamic limit, where phase transitions are characterized by a discontinuity in an order parameter or by nonanalytic behavior of derivatives of the free energy. As the Ising transition is a second-order phase transition, the free energy remains continuous in the thermodynamic limit, while higher derivatives become nonanalytic at the critical point. At finite sizes, however, the partition function is a finite sum of analytic functions, so the free energy and all its derivatives are analytic. Consequently, the singular features observed in the thermodynamic limit are rounded: Discontinuities become smooth crossovers, and divergences are replaced by finite-size peaks that sharpen and shift with increasing system size.

As mentioned in Sec.~\ref{sec:conjecture}, the fidelity susceptibility can be used as an indicator of quantum phase transitions. Thus, its finite-size version is a good candidate for locating the finite-size phase boundaries. For this purpose we make use of the fidelity susceptibility $\chi_F$ that occurs in the expansion of the finite-size fidelity \eqref{eq:fidelity}.

By analyzing the parameter-dependent fidelity susceptibility we find that the location of the finite-size phase boundary generally approaches that of the infinite-size phase transition asymptotically. The technique is applied in two directions, given that the phase diagram is two-dimensional: The shift in the $h$ direction is determined while keeping $\Delta$ constant and vice versa.

The ``remodeled'' phase boundaries for a system of $L=16$ spins are depicted in Fig.~\ref{fig:shifted-phase-diags}. We are aware that the fidelity susceptibility is not a reliable measure for tracking phase transitions near the floating phase, as it typically exhibits peak revivals that depend on the system size (cf. \cite{Yu2022}).

Some violations, such as those illustrated in Figs.~\ref{fig:PM-approach-quad}\textcolor{red}{(d)} and \ref{fig:mother_fail}\textcolor{red}{(c)}, remain inexplicable by the adjusted positions of the phase transitions. However, other instances, such as the quenches depicted in Figs.~\ref{fig:mother_fail}\textcolor{red}{(a)}, \ref{fig:violations-size-dep}\textcolor{red}{(a)}, and \ref{fig:violations-size-dep}\textcolor{red}{(b)}, can be accounted for by these revised phase transition locations.

\section{Overlaps in the Post-Quench Spectrum}
\label{sec:oe}
\begin{figure}
    \centering
    \begin{subfigure}{0.5\textwidth}
        \centering
        \begin{overpic}[width=0.75\textwidth]{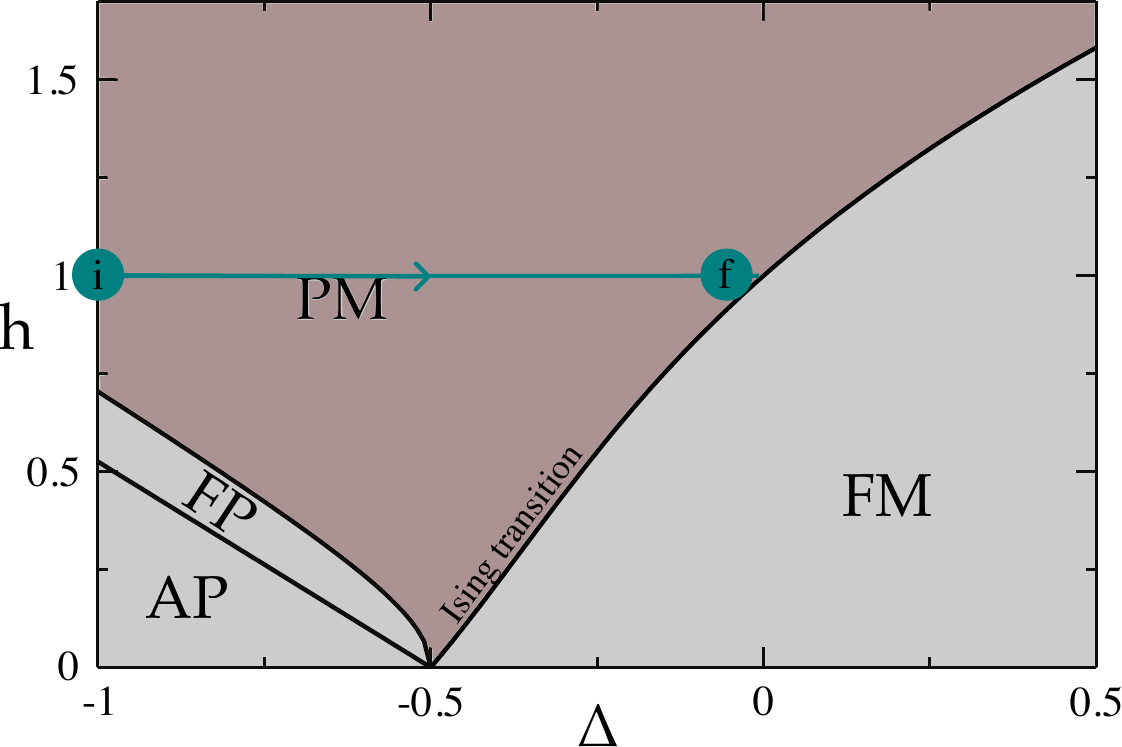}
        \put(-9,63){\textbf{(a)}}
        \end{overpic}
    \end{subfigure}

    \begin{subfigure}{0.5\textwidth}
        \centering
        \begin{overpic}[width=0.9\textwidth]{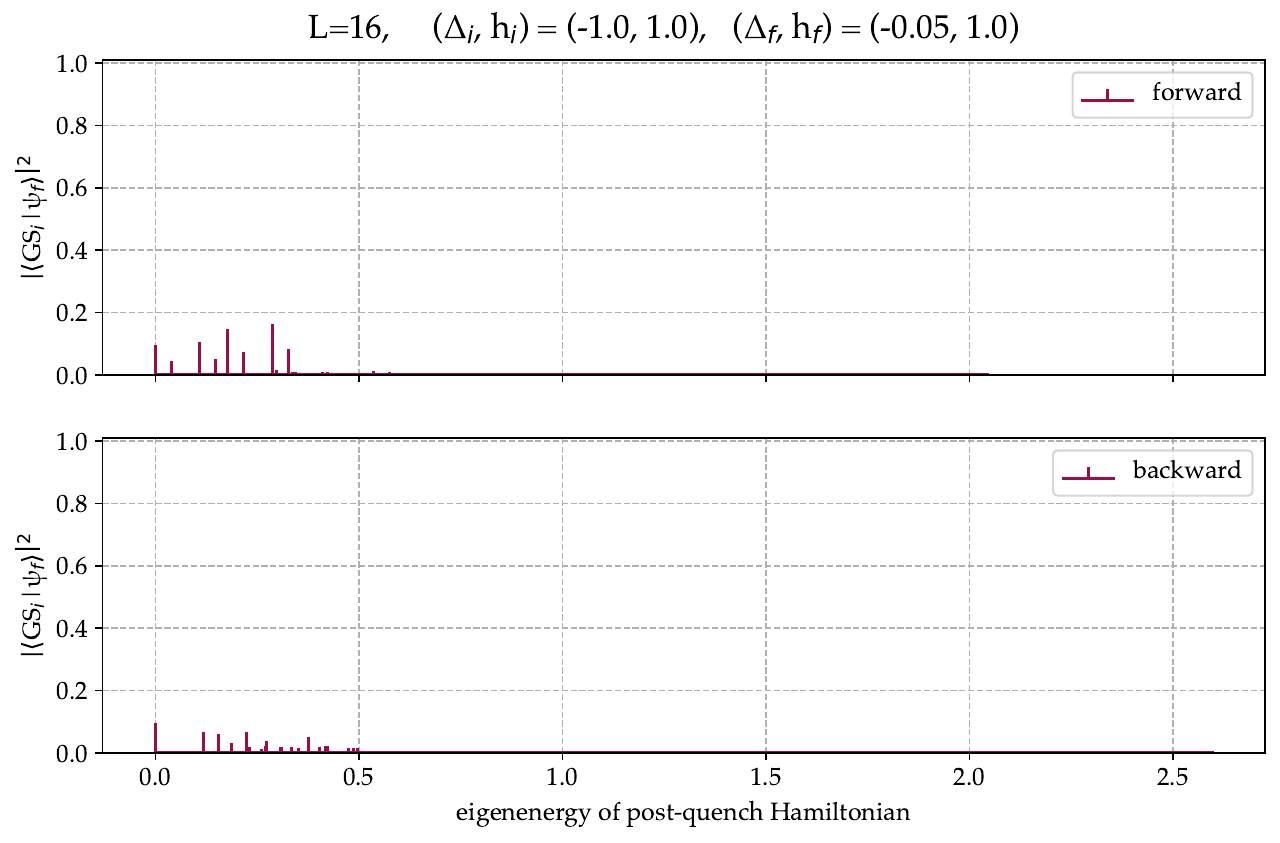}
        \put(-5,63){\textbf{(b)}}
        \end{overpic}
        % \captionsetup{labelfont={color=white}}
        % \caption{}
        \label{subfig:failed-failed}
    \end{subfigure}
    \caption{Quench within the paramagnetic phase showing ambiguous behavior.}
    \label{fig:failed-quench-again}
\end{figure}

For quenches within the same phase, the conjecture states that the largest overlap with the initial ground state should be that with the final ground state. Consequently, when one plots the overlaps against the postquench Hamiltonian spectrum, the largest peak should align with the lowest energy state. In Figs.~\ref{fig:PM}-\ref{fig:across_plots2}, we shift the postquench spectrum for this purpose such that the ground state energy is located at $0.0$ and all excitations energies are positive.

Figure \ref{fig:PM}\textcolor{red}{(a)} shows quenches within the paramagnetic phase, carried out along the depicted trajectories. The direction labeled ``forward'' denotes a quench from $i$ to $f$, while ``backward'' indicates a quench from $f$ to $i$. Quenches within the floating phase, antiphase, and ferromagnetic phase are presented below using the same labeling scheme.

The results in Figs.~\ref{fig:PM}-\ref{fig:across_plots2} in the Appendix were obtained employing exact diagonalization. They support the conjecture, as the largest peak consistently occurs at the ground state energy.

This rendering of the results allows for a reinvestigation of the quench depicted in Fig.~\ref{fig:mother_fail}\textcolor{red}{(c)}, which fails to support the conjecture as points near the Ising transition exhibit a larger overlap than the ground state overlap. A parameter combination for the quench starting at the same point as in Fig.~\ref{fig:mother_fail}\textcolor{red}{(c)} and ending in the darker region close to the phase boundary is used for the result illustrated in Fig.~\ref{fig:failed-quench-again}.

The result appears ambiguous: The backward quench supports the conjecture's statement, whereas the forward result does not. In the top panel of Fig.~\ref{fig:failed-quench-again}\textcolor{red}{(b)}, the largest peak is not located at the ground state energy but at higher energies in the spectrum. In fact, the heights of three peaks exceed that of the ground state overlap. A possible, although not certain explanation for this observation could be the finite-size shifted phase boundaries shown in Fig.~\ref{fig:shifted-phase-diags}.

To demonstrate that the above results are not a generic feature of any quenched system but, rather, provide support for the conjecture, we show some quenches across phase transitions in the Appendix. For such quenches, the largest peak is likely to occur at higher energy levels. Figures \ref{fig:across_plots-FMPM}, \ref{fig:across_plots-PM_FP}, and \ref{fig:across_plots2} show quenches across various phase boundaries of the ANNNI model that were carried out along the depicted trajectories.

In contrast to the quenches within the same magnetic phase, the maximum overlap does not align with the ground state energy. Exceptions to this observation arise for quenches between the antiphase and floating phase as can be seen in Fig.~\ref{fig:across_plots2} in the Appendix. The finite-size shifted phase diagrams in Fig.~\ref{fig:shifted-phase-diags} suggest that the antiphase is extended to significantly higher values of $h$ for the finite systems studied and thus provide a possible explanation for why the antiphase--floating phase transition is not traceable.

\section{Conclusion}
\label{sec:conclusion}
This work aimed to investigate the relationship between eigenstates of quenched Hamiltonians, specifically testing the validity of the conjecture formulated in Sec.~\ref{sec:conjecture}.

We conducted calculations in the TFIM and the ANNNI model. For the former, which is analytically solvable through Jordan-Wigner transformation and Bogoliubov rotation, we were able to establish a general analytical proof of the conjecture independent of specific quench points. For the ANNNI model, we could not attain a complete analytical proof because the model is not exactly solvable. Instead, we verified the conjecture in a special case where exact expressions for the pre- and postquench ground states are available. To extend the analysis beyond this special case, we employed numerical methods.

The numerical analysis comprises ED and the Lanczos method to compute the Hamiltonian's full spectrum or relevant parts of it. In the TFIM, numerical findings support the conjecture for all cases. Numerical results for the ANNNI model also mostly support the conjecture. While some violations could successfully be attributed to finite-size effects, for others whether they can be explained with finite-size shifted phase boundaries is not certain. Employing other methods for quantifying finite-size effects may offer a more comprehensive explanation for the observed violations.

Overall, the findings cautiously affirm the conjecture, suggesting it is an extension of the adiabatic theorem by Born and Fock\cite{Born1928} to maximally nonadiabatic changes. We acknowledge that counterexamples can be constructed, at least for systems without a continuous spectrum. However, whether the conjecture holds universally for all systems with a continuous spectrum in the thermodynamic limit or whether its validity must be restricted to a specific class of Hamiltonians or to quench end points that are relatively close to the quench start point remains unclear. Further investigation is needed to determine the precise conditions under which the conjecture holds. Nevertheless, our results strongly suggest that a significant class of Hamiltonians satisfies the conjecture, making it worthwhile to explore the conditions under which it remains valid.

\section{Acknowledgement}
We thank D. Schuricht for his valuable input, which contributed significantly to this paper. We acknowledge support from the Deutsche Forschungsgemeinschaft (DFG, German Research Foundation), Grant No. 499180199 (Project T1) via
FOR 5522.

\section{Data Availability}
The data that support the findings of this article are openly available \cite{zenodo}.
% \par\vspace{35\baselineskip}\noindent
\newpage

\section{Appendix}
\begin{figure}[h]
    \centering
    \begin{subfigure}{0.5\textwidth}
        \centering
        \begin{overpic}[width=0.75\textwidth]{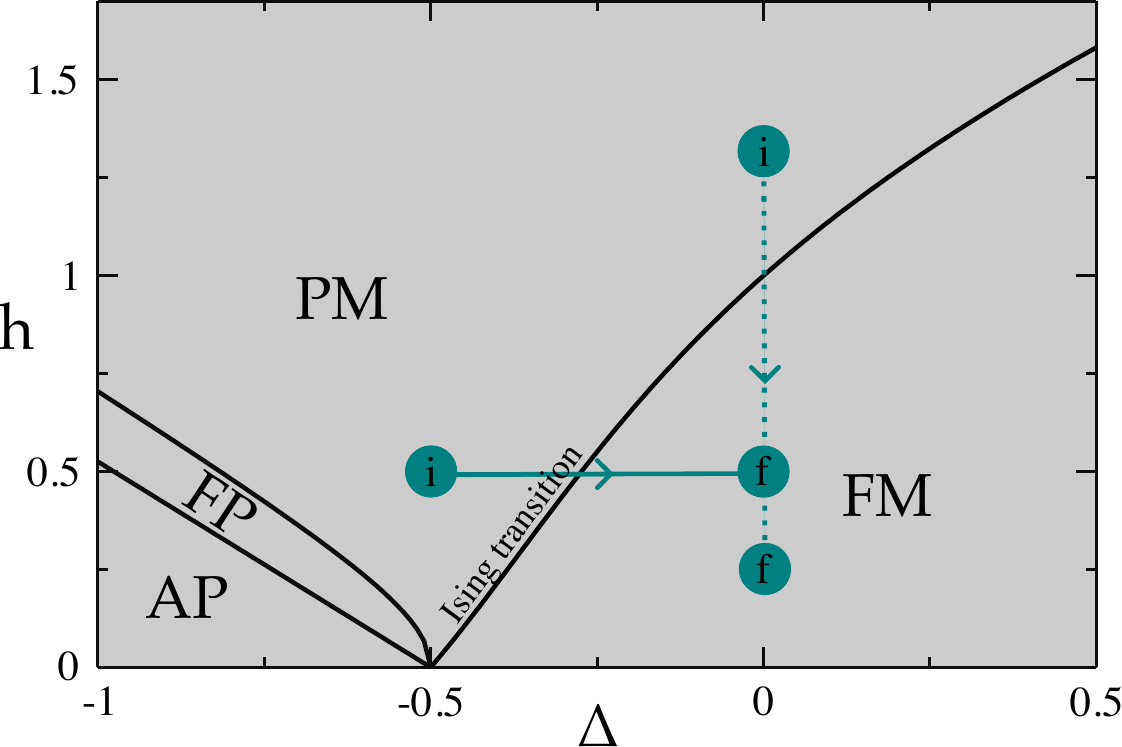}
        \end{overpic}
    \end{subfigure}

    \begin{subfigure}{0.5\textwidth}
        \centering
        \begin{overpic}[width=0.95\textwidth]{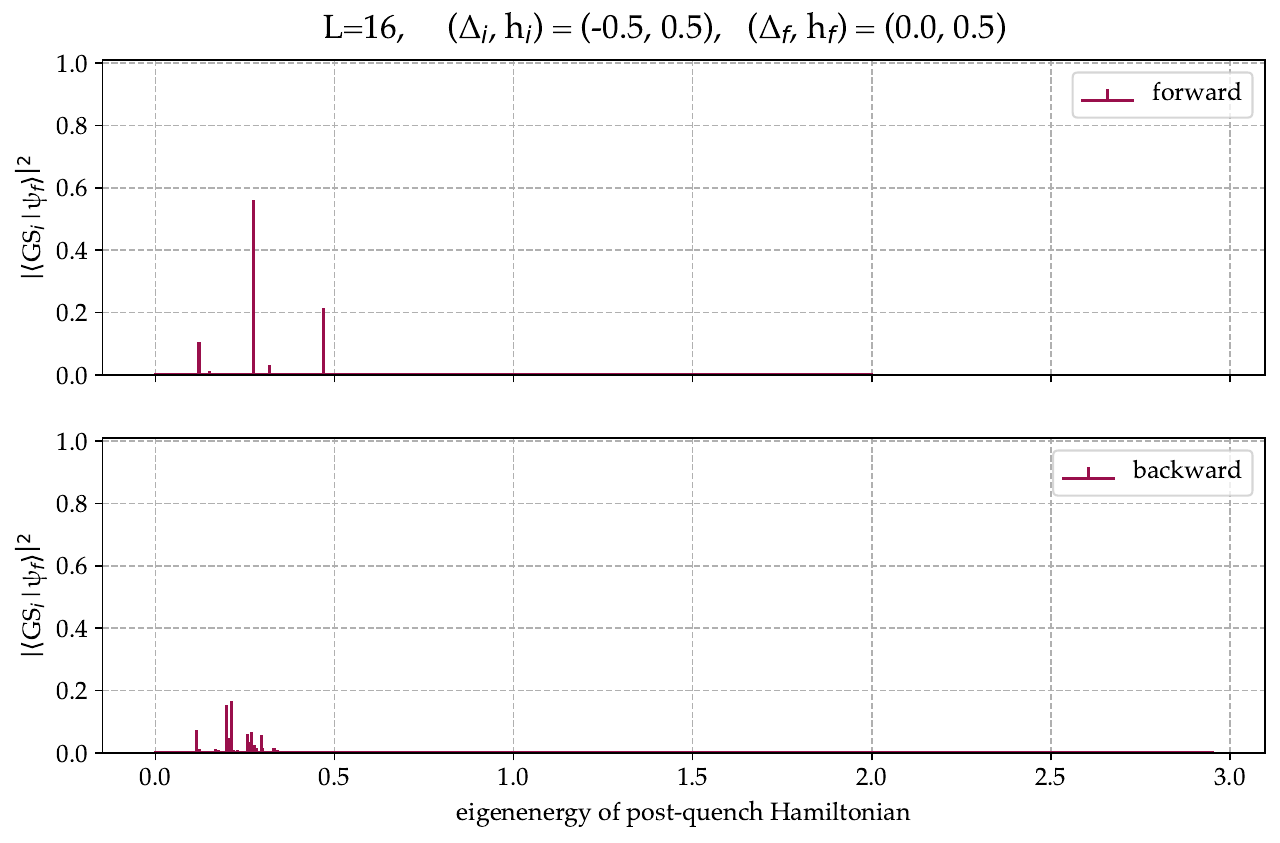}
        \put(-5,60){\textbf{(b)}}
        \end{overpic}
    \end{subfigure}

    \begin{subfigure}{0.5\textwidth}
        \centering
        \begin{overpic}[width=0.95\textwidth]{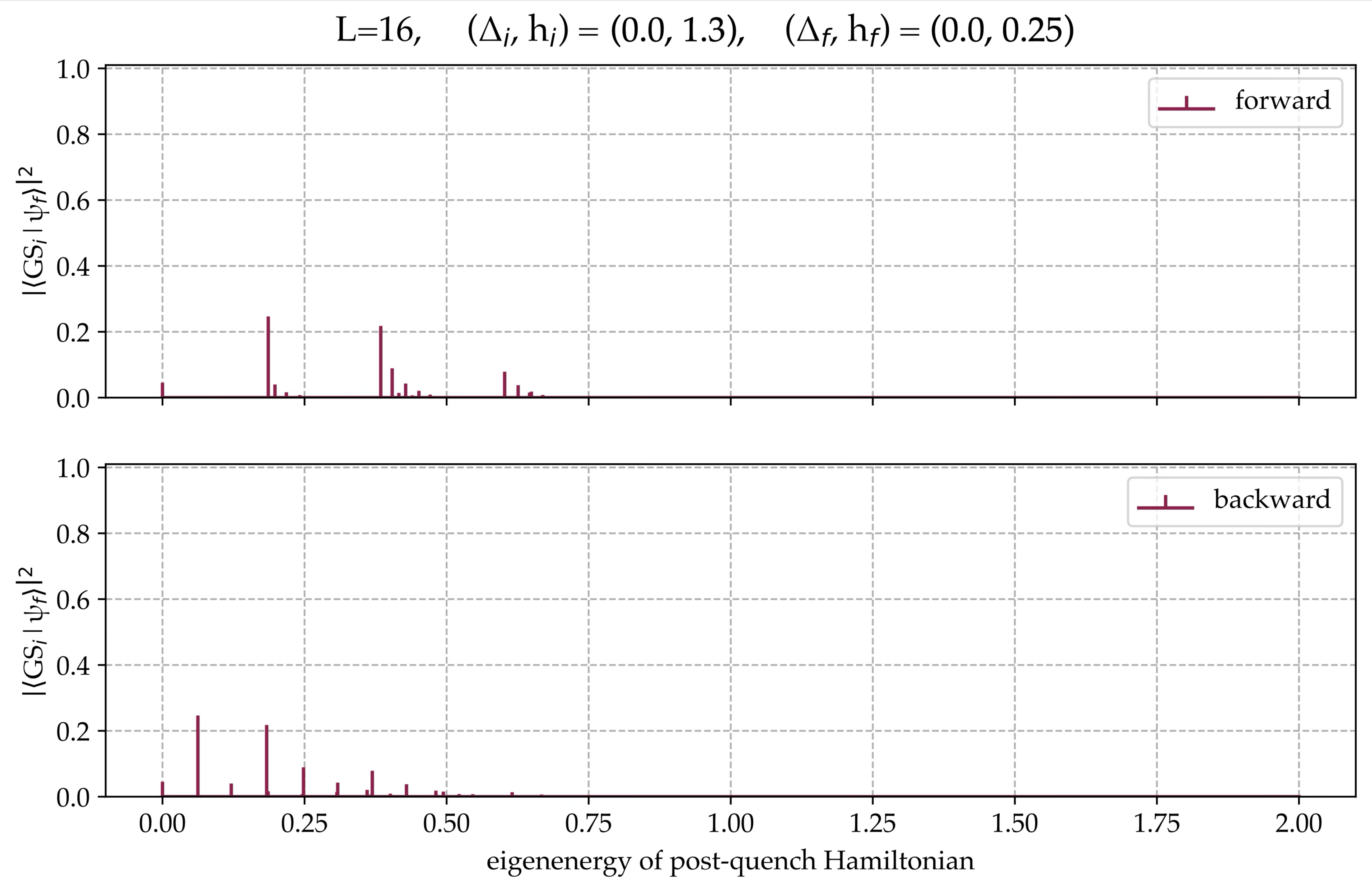}
        \put(-5,60){\textbf{(c)}}
        \end{overpic}
    \end{subfigure}
    \caption{Quenches across the paramagnetic--ferromagnetic phase transition. The phase diagram in (a) illustrates the quench trajectories. The quench direction from $i$ to $f$ is referred to as ``forward.'' Adapted from Karrasch and Schuricht \cite{Karrasch2013}.}
    \label{fig:across_plots-FMPM}
\end{figure}

\begin{figure}[h]
    \centering
    \begin{subfigure}{0.5\textwidth}
        \centering
        \begin{overpic}[width=0.75\textwidth]{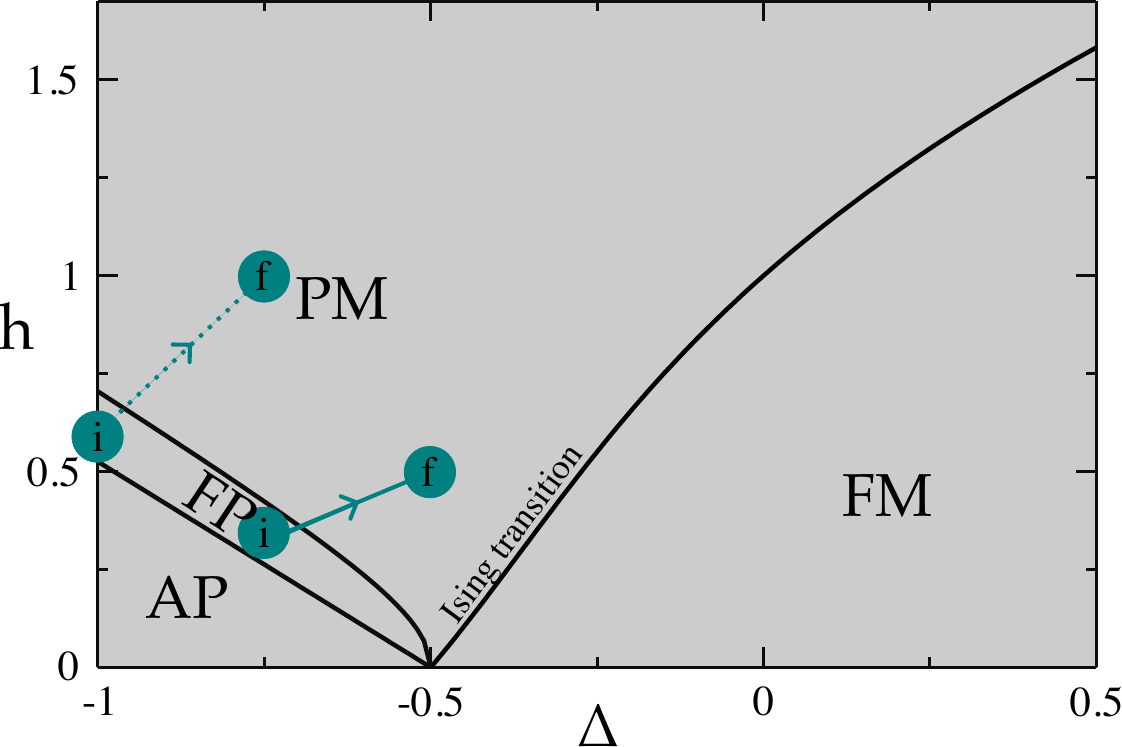}
        \end{overpic}
    \end{subfigure}

    \begin{subfigure}{0.5\textwidth}
        \centering
        \begin{overpic}[width=0.95\textwidth]{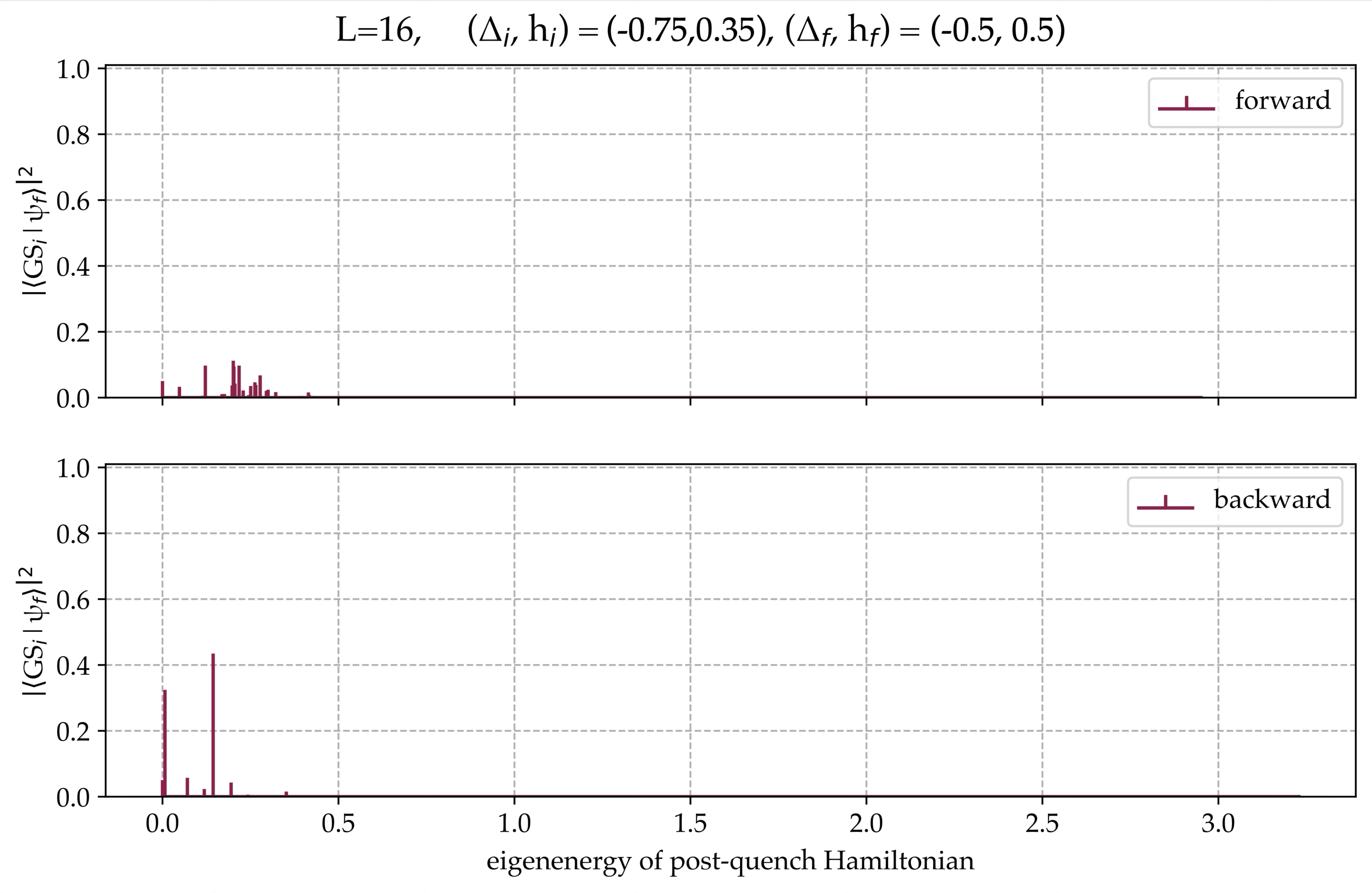}
        \put(-5,60){\textbf{(b)}}
        \end{overpic}
    \end{subfigure}

    \begin{subfigure}{0.5\textwidth}
        \centering
        \begin{overpic}[width=0.95\textwidth]{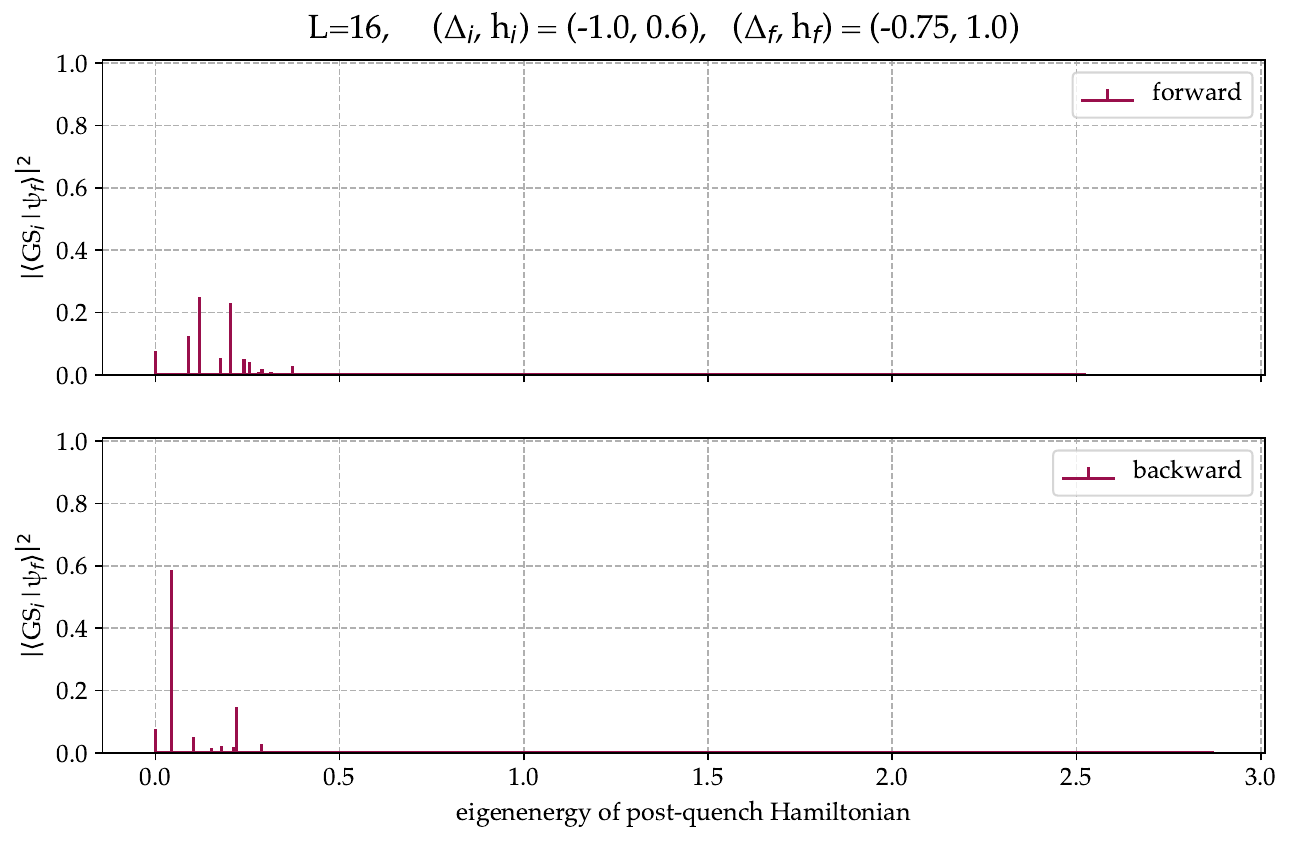}
        \put(-5,60){\textbf{(c)}}
        \end{overpic}
    \end{subfigure}
    \caption{Quenches across the floating phase--paramagnetic phase transition. The phase diagram in (a) illustrates the quench trajectories. The quench direction from $i$ to $f$ is referred to as ``forward.'' Adapted from Karrasch and Schuricht \cite{Karrasch2013}.}
    \label{fig:across_plots-PM_FP}
\end{figure}

\begin{figure}[h!]
    \centering
    \begin{subfigure}{0.5\textwidth}
        \centering
        \begin{overpic}[width=0.75\textwidth]{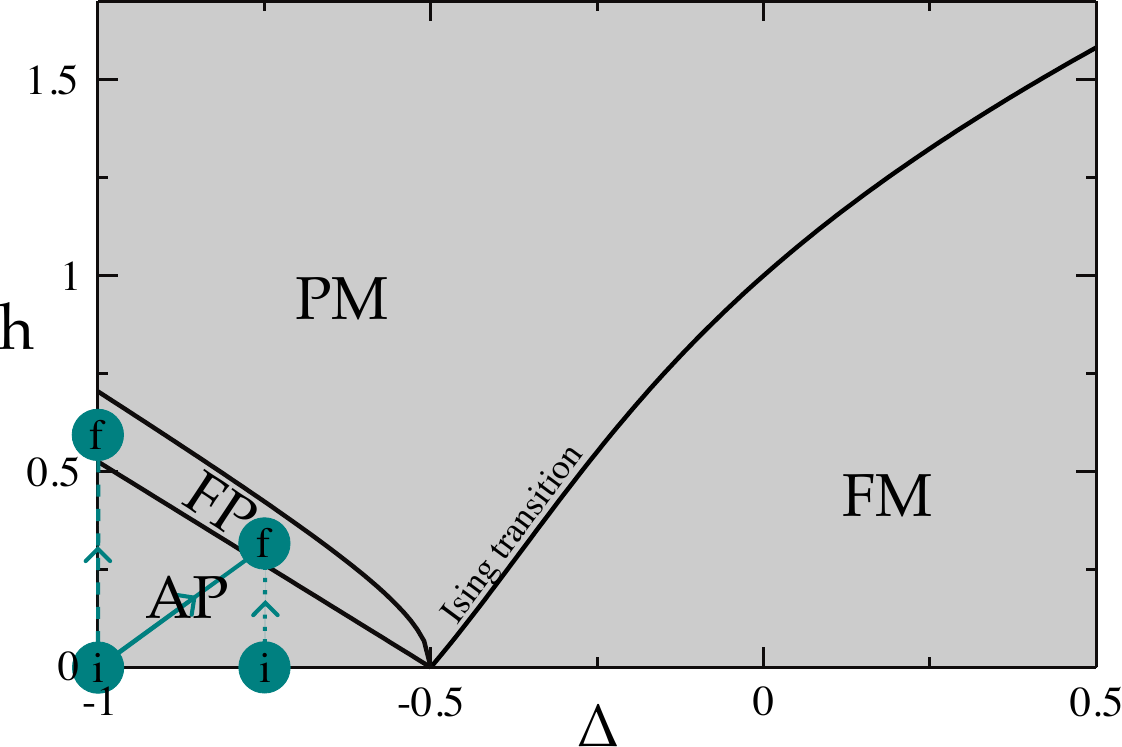}
        \end{overpic}
    \end{subfigure}

    \begin{subfigure}{0.5\textwidth}
        \centering
        \begin{overpic}[width=0.95\textwidth]{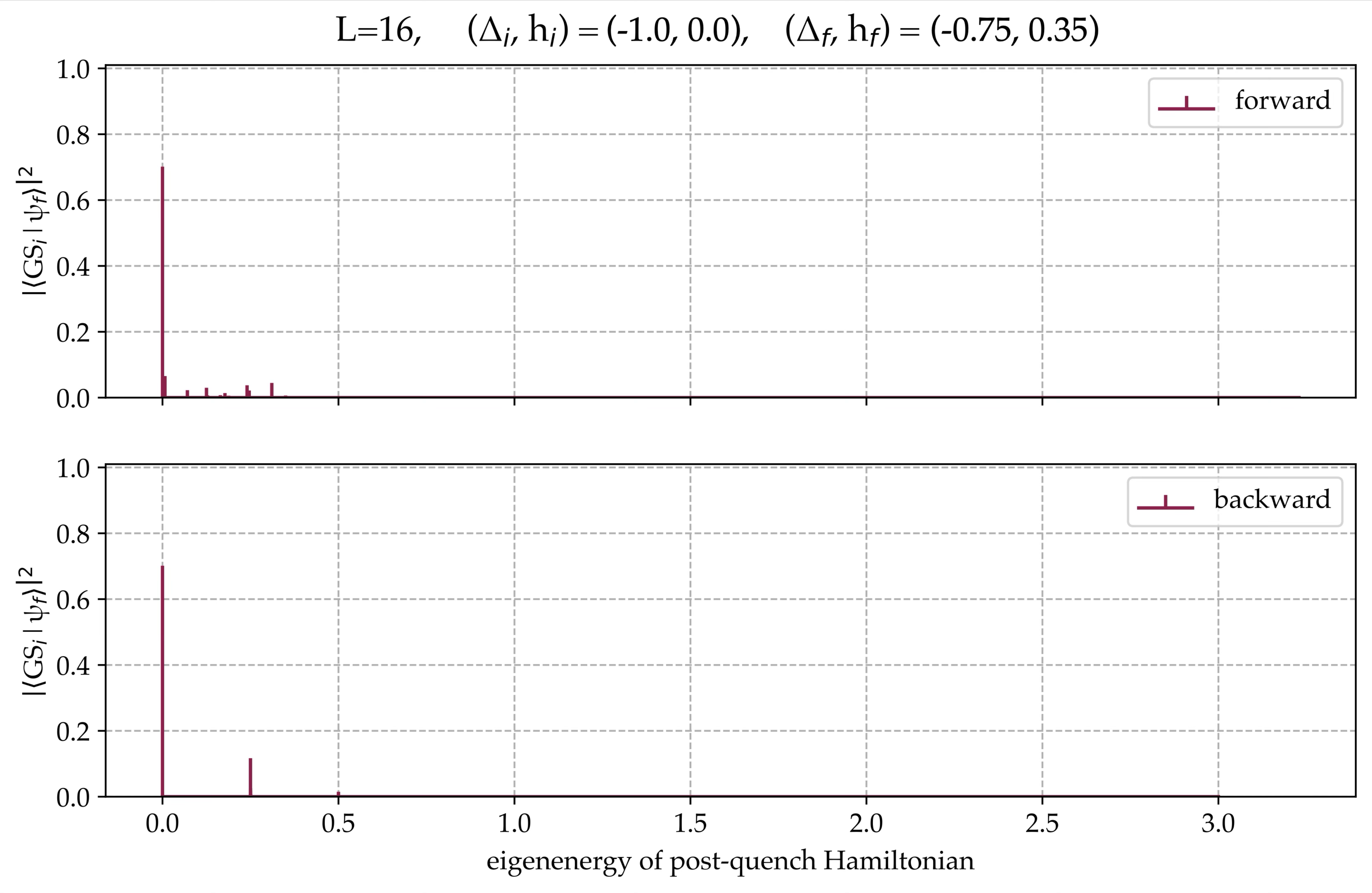}
        \put(-5,60){\textbf{(b)}}
        \end{overpic}
    \end{subfigure}

    \begin{subfigure}{0.5\textwidth}
        \centering
        \begin{overpic}[width=0.95\textwidth]{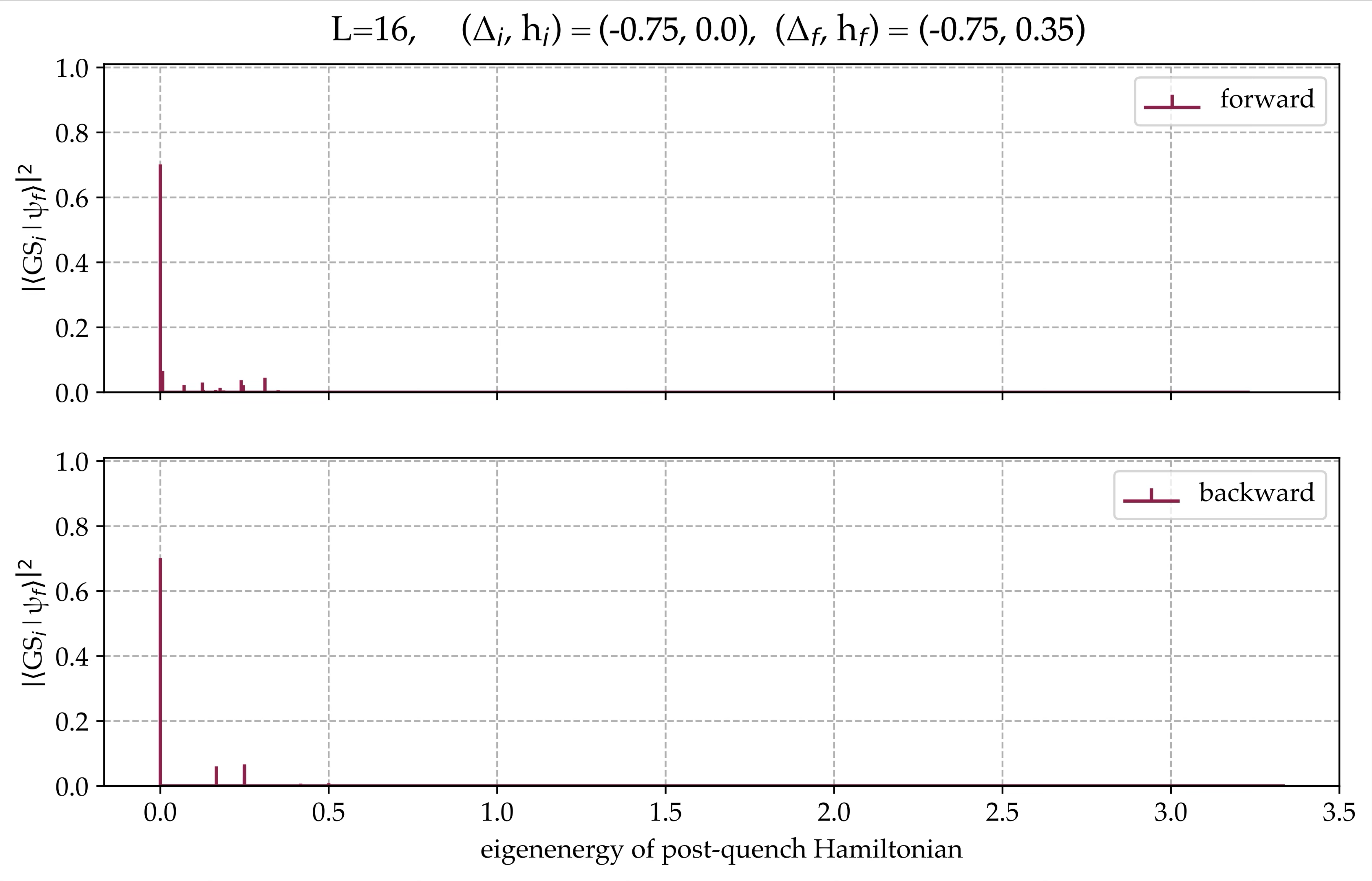}
        \put(-5,60){\textbf{(c)}}
        \end{overpic}
    \end{subfigure}

    \begin{subfigure}{0.5\textwidth}
        \centering
        \begin{overpic}[width=0.95\textwidth]{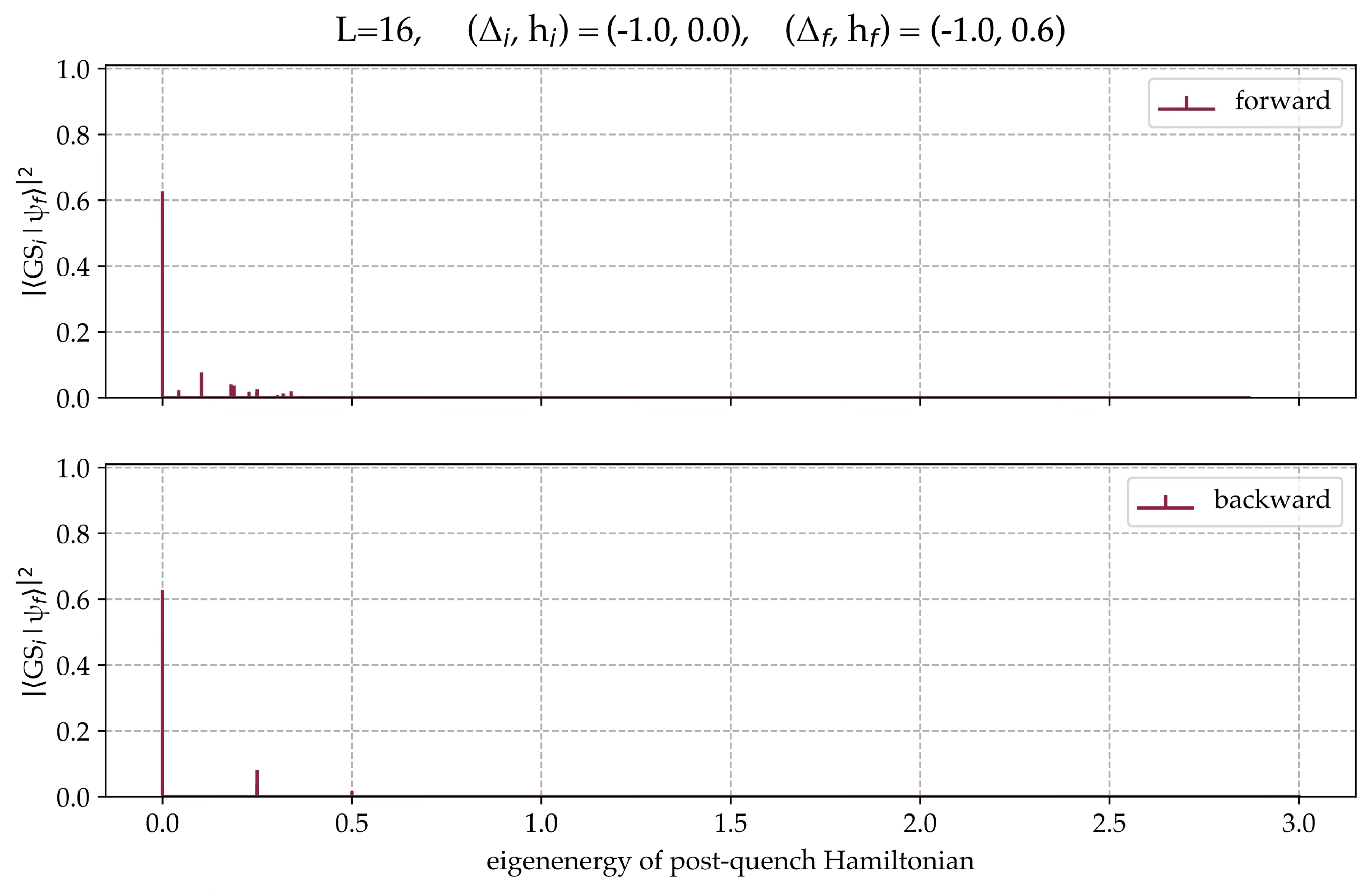}
        \put(-5,60){\textbf{(d)}}
        \end{overpic}
    \end{subfigure}
    \caption{Quenches across the floating phase--antiphase transition. The phase diagram in (a) illustrates the quench trajectories. Adapted from Karrasch and Schuricht~\cite{Karrasch2013}.}
    \label{fig:across_plots2}
\end{figure}

Figures \ref{fig:across_plots-FMPM}, \ref{fig:across_plots-PM_FP}, and \ref{fig:across_plots2} show quenches across various phase
boundaries of the ANNNI model that were carried out along
the depicted trajectories.
\newpage

\newpage

% \bibliographystyle{apsrev4-2}
% \bibliography{bib}


\clearpage
\end{document}